\documentclass[aps,prc,twocolumn,superscriptaddress,floatfix,longbibliography]{revtex4-2}

\usepackage{microtype}
\usepackage{graphicx}
\usepackage{amsmath,amssymb,bm}
\usepackage{xcolor}

\definecolor{rose}{RGB}{172,84,121}
\usepackage[colorlinks=true,linkcolor=rose,citecolor=rose,urlcolor=rose]{hyperref}

\begin{document}

\title{Diffusion-Based Point-Cloud Generation of Heavy-Ion Events}

\author{Rita Sadek}
\affiliation{Nuclear Science Division, Lawrence Berkeley National Laboratory, Berkeley, CA 94720, USA}

\author{Vinicius Mikuni}
\affiliation{Nagoya University, Kobayashi-Maskawa Institute, Aichi 464-8602, Japan}
\author{Mateusz Ploskon}
\affiliation{Nuclear Science Division, Lawrence Berkeley National Laboratory, Berkeley, CA 94720, USA}

\begin{abstract} 
      Heavy-ion collisions produce final states with thousands to tens of thousands of particles, making their simulation among the most computationally intensive tasks in high-energy nuclear physics. 
      We present a fast, high-fidelity generative model for heavy-ion events based on a score-driven diffusion process and the Point-Edge Transformer architecture within the OmniLearn framework. 
      A two-stage training strategy is performed: Stage-1 training on lower-multiplicity O-O collisions allowing the model to learn a stable event and particles representation, followed by fine-tuning on challenging high-multiplicity Pb-Pb collisions. We benchmark the generator with a broad set of closure checks, including agreement of event- and particle-level observables in one and two dimensions, flow consistency reconstructed from the generated particles, end-to-end jet finding with FastJet including key jet and substructure observables, and a classifier-based application to quantify the sample fidelity. The results are promising, showing that a compact generative model can produce realistic, high-multiplicity heavy-ion events, at a level that makes local-scale generation for heavy-ion collisions at high energies a practical goal. 
\end{abstract}

\maketitle


\section{Introduction}
Heavy-ion collisions at high energies produce high multiplicity final states, in which jet and multi-particle observables probe both hard and collective dynamics of the medium. As measurements become more differential and precision-driven, jet analyses and correlation studies face a persistent challenge: the dominant backgrounds are largely combinatorial and must be estimated with high statistical precision across finely binned event selections. 
A standard tool in this context is the mixed-event technique, where particles from different events are combined to model the uncorrelated background and instrumental effects in jet, resonance, and correlation measurements. In practice, the mixed-event quality can drive leading systematic uncertainties and requires large, well matched event pools over centrality and event-shape selections, thus becoming both storage and computationally expensive, especially as analyses move towards higher precision and more differential measurements with the future High-Luminosity Large Hadron Collider \cite{hllhc}.

In this paper, we explore a machine learning method based on diffusion generative model application: rather than relying exclusively on storing and repeatedly re-sampling high multiplicity events, we aim to generate realistic events on demand while preserving both global event characteristics and the structure of particle-level kinematics. 
For mixed-event workflows, this is a stringent requirement: the generator must simultaneously reproduce single-particle spectra and multi-particle correlations, maintain global azimuthal structure associated with collective flow, and remain faithful after end-to-end jet finding and substructure measurements. 
We study heavy-ion event generation with a score-based Point Edge Transformer (PET) within the OmniLearn framework \cite{omnilearn}: a transformer-style diffusion generative model that treats the final state as a variable-length set of particle “tokens”, conditioned on event-level information. The model is thus trained to generate both event features and per-particle kinematics in a single pipeline, enabling direct use of generated particles as inputs to standard reconstruction, jet finding, and mixed-event construction chains.

This paper is organized as follows. Section 2 introduces the generator used in this work, including both the event-level diffusion branch (ResNet) and the particle-level diffusion (PET). Section 3 describes the simulated datasets, the event- and particle-level feature construction used for training, and the two-stage training strategy. Section 4 presents detailed results for O-O collisions, establishing baseline closure for event structure, flow-related observables, and jet reconstruction. Section 5 discusses fine-tuning on Pb-Pb collisions, and the additional constraints needed to maintain global coherence in a high-multiplicity challenging regime. Finally, Section 6 summarizes the conclusions and outlines the next steps. 

\section{Heavy-ion event generation with a point cloud-based diffusion model}

Our goal is to generate full heavy-ion events at the level needed for downstream jet and correlation analyses, which thus implies the following requirements: (i) a set of global event properties (centrality and geometry-related information, and global kinematic summaries) and (ii) a variable-length set of final-state particles with per-particle kinematics. We implement this as a conditional, score-based \cite{score_sde} generative model within the OmniLearn framework, where a diffusion process is learned for both event-level and particle-level representations. The model architecture follows the design introduced in Ref.~\cite{omnilearn_eic} and is illustrated in Figure 1.

\begin{figure}[!htb]
    \centering
    \includegraphics[width=1.\linewidth]{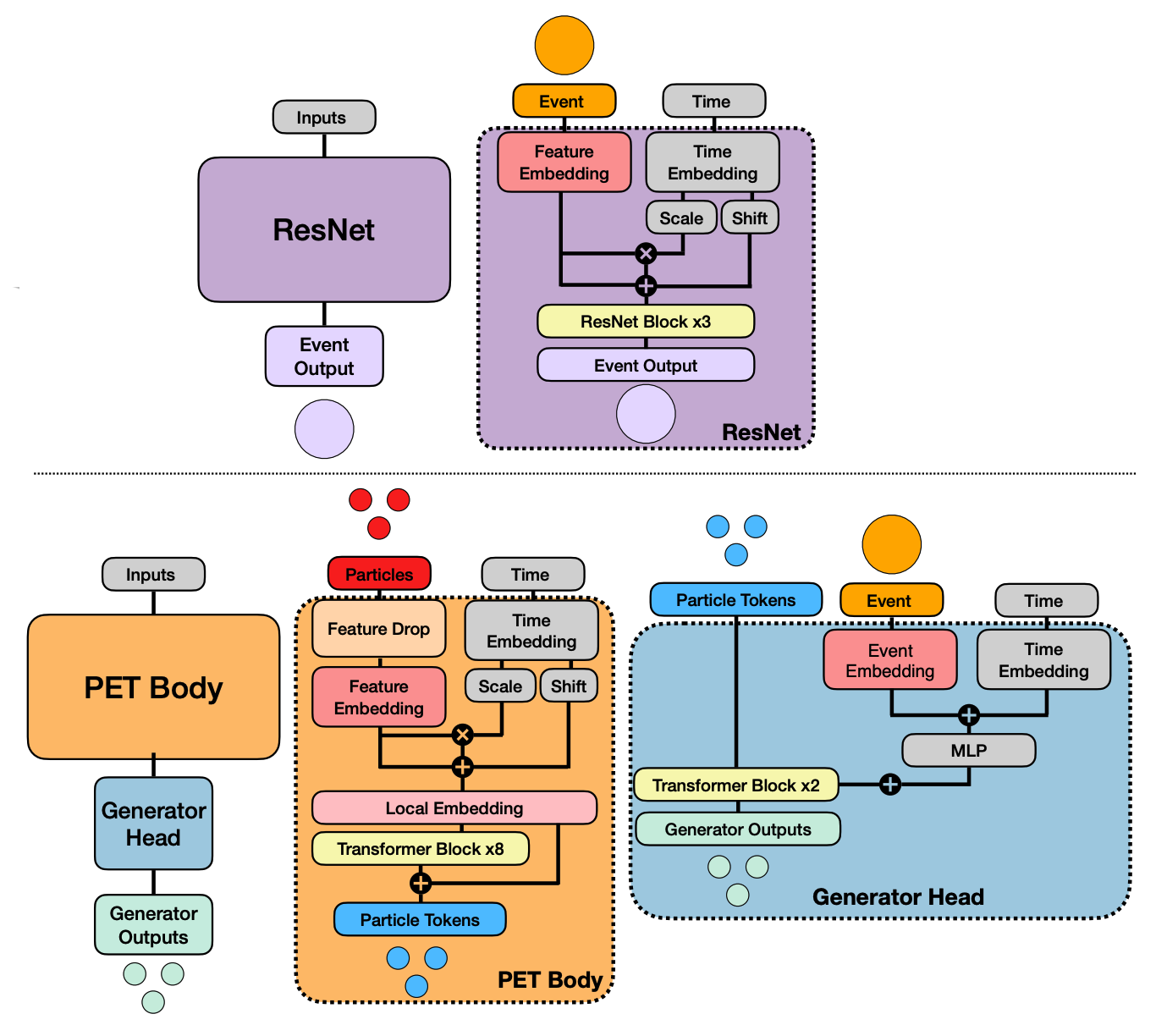}
    \caption{OmniLearn model architecture, showing the detailed main blocks used in this paper: ResNet followed by the PET body \cite{omnilearn_eic}.}
    \label{fig:ResNet_PET}
\end{figure}

\subsection{Conditional score-based formulation}

Each event is modeled as a conditional two-stage generation problem with (i) an event-level feature vector $e \in \mathbb{R}^{J}$ and (ii) a variable-length particle set $\{ {x_i} \} ^{N} _{i=1}$, $x_i \in \mathbb{R}^{ F}$. With $N$ reflecting the event multiplicity, thus varying event-by-event, we store particles in a fixed-size tensor of length $N_{\max}$ padding and a binary mask $m\in \{ 0,1\} ^{N_{\max}}$ to indicate valid particles. Generation proceeds in the following two steps: we first sample an event vector $e_0$, and then sample the particle set $x_0$ conditioned on $e_0$ (during training $e$ is taken from data; during generation $e=e_0$). This will be further discussed in the next subsection. 

Both steps are implemented with a diffusion model: a generative model that starts from Gaussian noise and iteratively de-noises. Given a clean target $z_0$ (where $z$ denotes either $e$ or $x$), a forward noising process is defined at a continuous diffusion time $t\in (0,1)$: 
\begin{equation*}
    z_t = \alpha(t)z_0 + \sigma(t)\varepsilon,\qquad \varepsilon\sim\mathcal{N}(0,I)
\end{equation*}
where $\alpha(t)$ and $\sigma(t)$ define the noise schedule and satisfy $\alpha^2 (t) +\sigma^2 (t)=1$ with $\alpha(t) = \cos(0.5\pi t)$ and $\sigma(t) = \sin(0.5\pi t)$ .
The learning task is then to approximate the reverse transformation $z_t\mapsto z_0$ for all noise levels $t$, conditional on the available information. The forward process destroys structure by adding noise; while the reverse model learns to recover structure by de-noising. Rather than predicting the noise $\varepsilon$ directly, we use the $v$-parameterization \cite{vparam}, commonly used in OmniLearn applications, where the network is trained to predict:
\begin{equation*}
    v(z_t,t,\cdot) = \alpha(t)\varepsilon - \sigma(t)z_0 
\end{equation*}
Concretely, the event branch takes $(e_t,t)$ as input and predicts $v_e$, while the particle branch takes $(x_t,t, e, m)$ and predicts $v_x$. Thus, from a prediction $v_\theta$,  we can obtain a de-noised estimate:
\begin{equation*}
    \hat z_0=\alpha(t) z_t - \sigma(t) v_\theta(z_t,t,\cdot)
\end{equation*}
where the training minimizes a mean-squared error between the predicted and the target velocities.

\subsection{Two-branch architecture: event and particle generator}

To implement the diffusion de-noisers, the two-branch architecture is used with separate networks for the event-level vector and the per-event particle set, as displayed in Figure 1. 

\textbf{Event branch:} The event de-noiser is a ResNet MLP that takes the noised event vector $e_t$ and the diffusion time $t$. The time embedding is injected through feature-wise scale/shift modulation, and the network predicts the event $v$-target $v_e$. During generation, this branch is used to sample an event vector $e_0$ starting from Gaussian noise by running the reverse diffusion sampler.

\textbf{Particle branch:} The particle de-noiser is based on a PET body consisting of stacked transformer blocks with a locality bias, followed by a conditional generator head. It takes the noised particle tensor $x_t$, the diffusion time $t$, the event-level conditioning vector $e$, and the padding mask $m$, and predicts the particle $v$-target $v_x$. The mask prevents padded entries from contributing to the computation and to the loss, enabling stable learning for variable multiplicity. 

With this two-branch design, event generation and particle generation are performed sequentially at inference time: we first sample $e_0$ with the event branch, which also determines the particle multiplicity to be generated, then sample the particle set $x_0$ with the particle branch conditioned on $e_0$. 

\section{Data, model and training strategy}

All results in this paper are obtained with Pythia8 Angantyr \cite{angantyr, pythia8} heavy-ion simulations for both O-O and Pb-Pb collisions at $\sqrt{s_{\mathrm{NN}}}=5.36\ \mathrm{TeV}$. Events are stored in HDF5 format with a fixed structure consisting of (i) an event-level array of shape $(N_{\mathrm{evt}}, 11)$, and (ii) a particle-level array of shape $(N_{\mathrm{evt}}, N_{\max}, 6)$. The same format is used for the training, testing and validation datasets, and generated samples are written in the identical structure to enable closure studies.

\subsection{Training splits and multiplicity regimes}

We use explicit train/test/validation splits to separate optimization from physics validation. For O-O collisions, we train on $10^6$ events and reserve $3\times10^5$ events for test and $10^5$ events for validation, which is used for the performance plots shown in Section 4. The O-O samples have a characteristic maximum multiplicity of $\mathcal{O}(10^3)$ particles per event.

For Pb-Pb collisions, the multiplicity increases by roughly an order of magnitude in the samples considered here, reaching $\mathcal{O}(10^4)$ particles per event. This large multiplicity gap motivates the two-stage training strategy described below in Section 3.3. The statistics used for this fine-tuning stage is approximately one-tenth the size of the O-O collisions Stage-1 sample size: we train on $10^5$ Pb-Pb events and reserve $3\times10^4$ for testing and $4\times10^3$ for validation.

\subsection{Event and particle feature vectors}

In our datasets, we choose the event-level vector to contain the following 11 scalars: 
\begin{equation*}
\begin{aligned}
e = (&\langle p_\mathrm{T}\rangle,\ \sum p_\mathrm{T},\ Q_x,\ Q_y,\ \psi_{\rm EP}, b,\\
     &\ N_{\rm coll},\ N_{\rm part},\  \langle \eta\rangle,\ \langle \phi\rangle,\ N_{\rm particles})
\end{aligned}
\end{equation*}
Here, the mean quantities are explicitly per-event averages over the final-state particles $N$:
\begin{equation*}
    \langle p_\mathrm{T}\rangle=\frac{1}{N}\sum_{i=1}^{N} p_{\mathrm{T},i} , \ \langle \eta\rangle=\frac{1}{N}\sum_{i=1}^{N} \eta_{i}, \ \langle \phi \rangle=\frac{1}{N}\sum_{i=1}^{N} \phi_{i}
\end{equation*}
The flow-vector components $(Q_x,Q_y)$ stored at event-level are the second-harmonic transverse flow vectors: 
\begin{equation*}
    Q_x=\sum_{i=1}^{N} p_{\mathrm{T},i} \cos(2\phi_i), \ Q_y=\sum_{i=1}^{N} p_{\mathrm{T},i} \sin(2\phi_i), 
\end{equation*}
The event-plane angle $\psi_2$ is calculated with the following: $\psi_2=\tfrac{1}{2}\mathrm{atan2}(Q_y,Q_x)$. 

The remaining components are generator-level geometry quantities describing the impact parameter $b$, the Glauber-model \cite{glauber} number of nucleon–nucleon collisions $ N_{\rm coll}$, the number of participants $N_{\rm part}$, and the final-state multiplicity $N_{\rm particles}$. 

Each particle token is represented by the 6-dimensional feature vector: 
\begin{equation*}
\mathbf{x}_i = \big(\phi^{\rm rel}_i,\ p{_\mathrm{T}}^{\rm rel}_i ,\ \eta_i^{\rm rel},\ p_{\mathrm{T}i} ,\ \eta_i,\ \phi_i\big) 
\end{equation*}
The relative coordinates are defined with respect to the event-level reference mean values, and the $p_{\mathrm{T},i}^{\rm rel}$ is defined as the following: $p_{\mathrm{T},i}^{\rm rel} = \log \left(\frac{p_{\mathrm{T},i}}{\langle p_\mathrm{T}\rangle}\right)$.

This approach allows the model to learn both the absolute kinematics and the event-normalized structure, stabilizing the learning process across events with varying multiplicity. All event-level and particle-level features are standardized prior to training by subtracting the training-set mean and dividing by the standard deviation, computed independently for each component. The multiplicity is normalized separately and mapped back to integer values during generation. The same preprocessing constants are applied at generation time to revert the model output to physical units. 

\subsection{Implementation, compute constraints, and two-stage strategy}

In the \textbf{particle branch}, locality is introduced through the PET neighborhood construction, which operates on a geometry embedding rather than the full particle feature vector. In our implementation, the geometry coordinates passed to the PET body are the two relative particle-token features $g_i = (\phi^{\mathrm{rel}}_i,\ p_{\mathrm{T}i} ^{\mathrm{rel}})$, while the full six-dimensional token remains available to the network for de-noising. This design choice applies only to the neighborhood definition (locality bias) and was selected empirically in our setup as it led to stable optimization and an improved learned structure.

Training follows the conditional diffusion objective described in Section 2, using the same noise schedule and $v$-parameterization for both event and particle branches. The dominant practical constraint is GPU memory in the particle branch. PET uses transformer blocks that model correlations within the particle set by computing pairwise interaction weights between particles. Concretely, for a padded particle length $N_{\max}$, the model constructs a table of interaction scores between all particle pairs, replicated over a number $H$ of  parallel heads (independent attention subspaces). This yields a tensor of scores with shape $[B, H, N_{\max}, N_{\max}]$, where $B$ denotes the batch size. These scores are then converted into normalized weights by applying a softmax (a normalization that aligns scores to non-negative weights that sum to one over the neighbor index), and the resulting weights are used to aggregate information across particles. This construction implies memory that grows approximately quadratically with $N_{\max}$. In addition, the training step further increases the memory demand as it must retain the computational graph needed to compute gradients for the back-propagation purpose. Consequently, increasing the maximum multiplicity rapidly increases the memory footprint: the regime $N_{\max}\sim\mathcal{O}(10^3)$ (O-O) is tractable, while $N_{\max}\sim\mathcal{O}(10^4)$ (Pb-Pb) becomes a limiting factor, forcing much smaller batch sizes and motivating a staged training procedure.

Thus, we adopt the two-stage strategy aimed at transferring a stable event-particle representation across multiplicity regimes. \textbf{Stage~1 (O-O training):} we train the full generator (event and particle branches) on O-O events to learn robust single-particle and multi-particle structure in a moderate-multiplicity environment, establishing baseline closure for event-level distributions, particle-level kinematics, and flow-related observables reconstructed from particles. \textbf{Stage~2 (Pb-Pb fine-tuning):} we initialize the weights from the O-O checkpoint and fine-tune on Pb-Pb events, thus adapting to a new higher-multiplicity regime.

\section{Results for O-O collisions}

\subsection{Model configuration and checkpoint selection}

The O-O results presented in this section are obtained from the Stage-1 model trained on the Perlmutter supercomputer \cite{perlmutter} in data-parallel mode using Horovod \cite{horovod} across 5 GPU nodes (20 GPUs total). Training is performed with the OmniLearn TensorFlow/Keras implementation \cite{tensorflow}. We use a per-rank (local) batch size of 64 and train for 75 epochs. Optimization uses the Lion optimizer \cite{lion} with parameters $\beta_1=0.95$ and $\beta_2=0.99$. The learning rate follows a cosine-decay schedule starting from $3\times10^{-5}$ with a warmup phase corresponding to three epochs. The full model contains 2.15M trainable parameters in total (event branch + PET + particle generator head).

Throughout training, the best-performing weights are tracked on the held-out validation loss and automatically saved; all plots shown in this section are produced using the best saved checkpoint (lowest validation loss) rather than the final epoch.
In early development we tested strict early-stopping criteria based on rapid validation-loss stabilization; this typically terminated training much earlier but led to degraded closure in downstream physics observables. A more detailed quantification of the quality between different models is described in Section 4.5. 
In the following, we thus retain a fixed-length training while saving the best checkpoint, which allows the optimizer to reach an improved minima, not captured by early plateau-based stopping.

\begin{figure*}[p]
    \centering
    \includegraphics[width=1\linewidth]{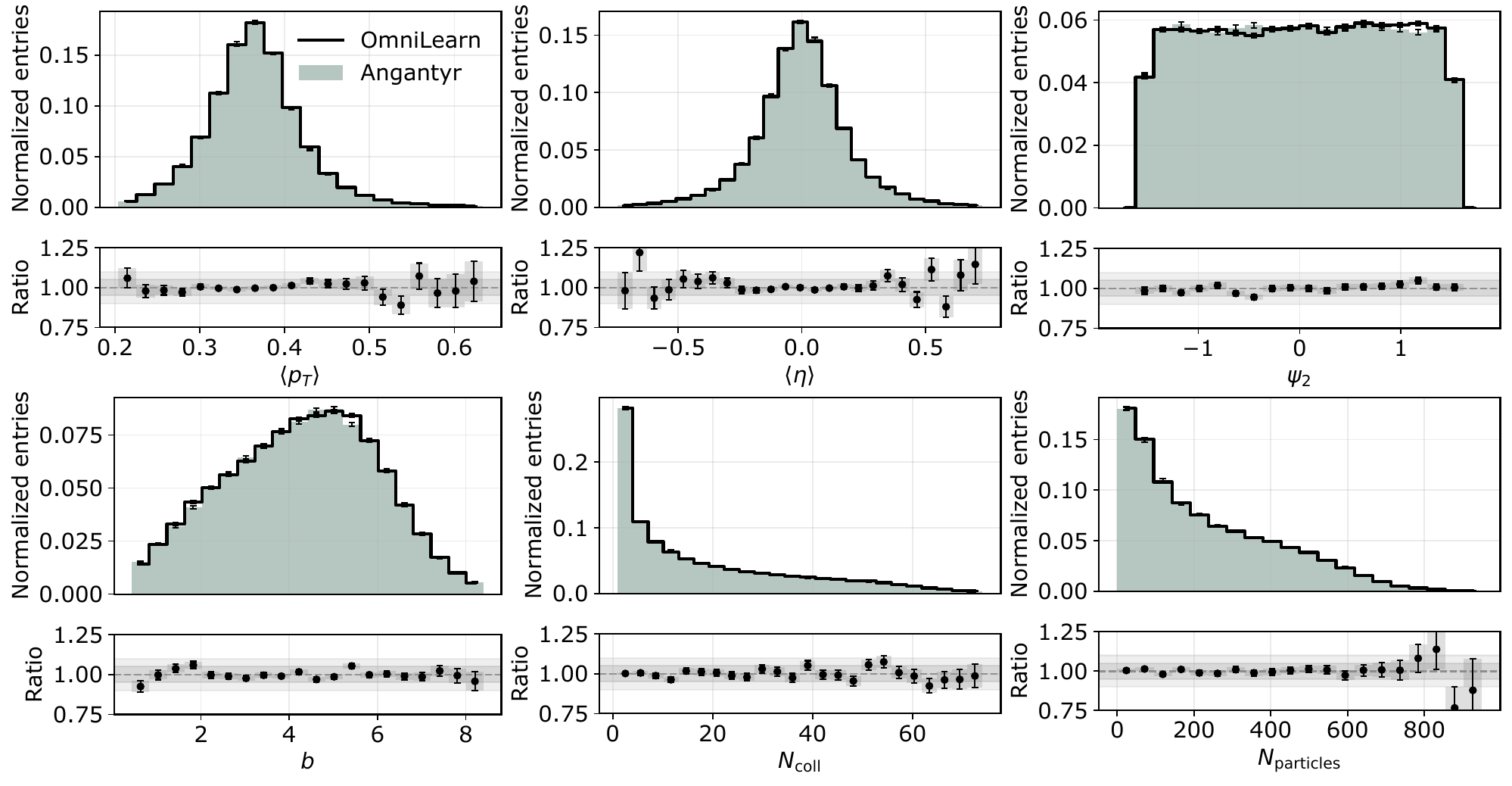}
    \caption{Event-level one-dimensional closure tests for the Stage-1 O-O generator. Shown are representative global event properties used for particles conditioning and generated by the event branch: $\langle p_\mathrm{T}\rangle$, $\langle \eta\rangle$, the second-harmonic event-plane angle $\psi_2$, the impact parameter $b$, the number of nucleon–nucleon collisions $N_{\rm coll}$, and the event multiplicity $N_{\rm particles}$. All distributions are normalized to unit area. The lower panels show the ratio of generated-to-validation distributions with statistical uncertainties from bin counts. The gray bands delimit $\pm5\%$ and $\pm10\%$ around unity.} 
    \label{fig:figOO_event1D}
\end{figure*}

\begin{figure*}[p]
    \centering
    \includegraphics[width=1\linewidth]{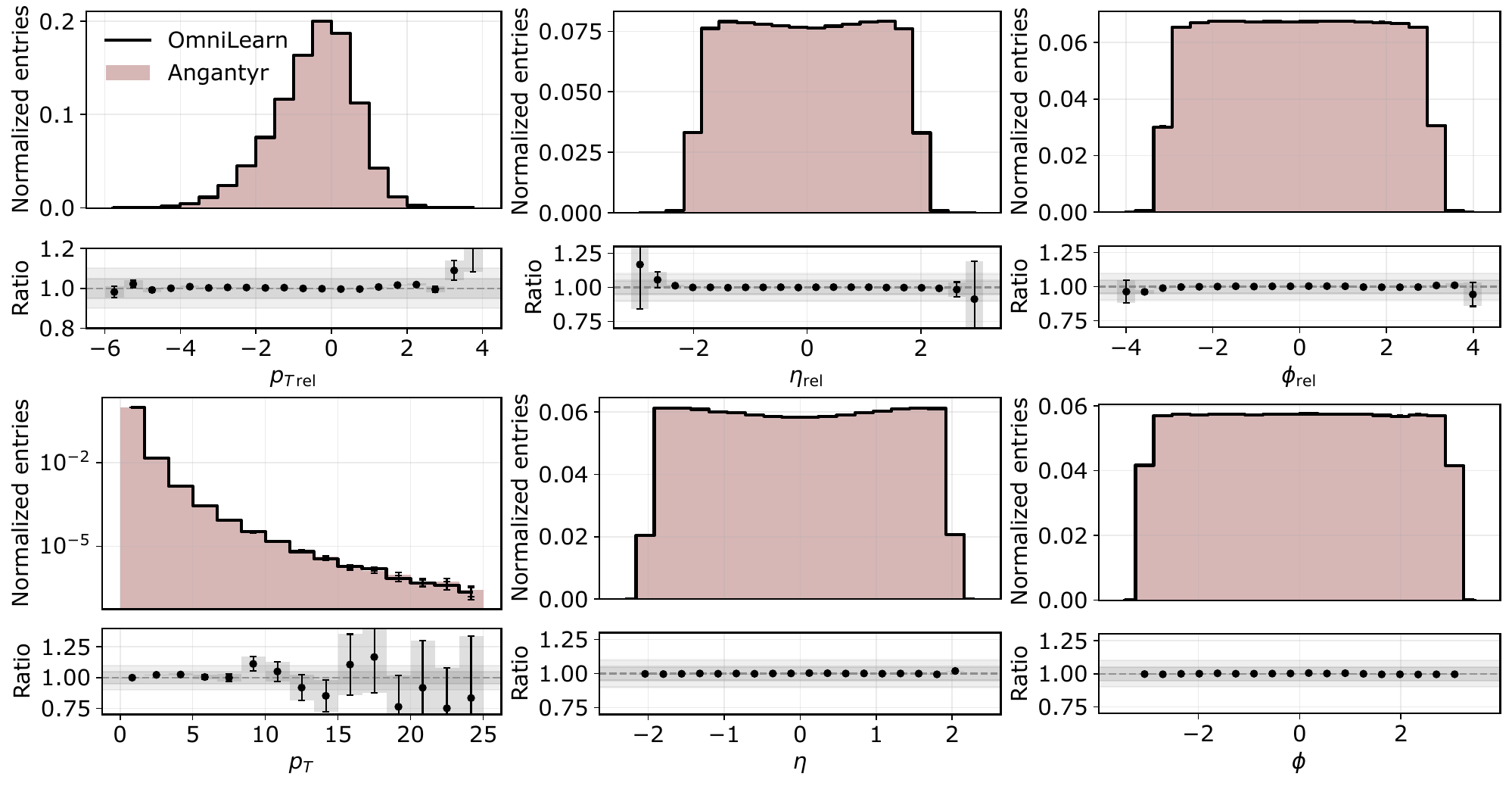} 
    \caption{Particle-level one-dimensional closure tests for the Stage-1 O-O generator.
Distributions are shown for the six-dimensional particles representation, including the relative variables $(p_\mathrm{T}^{\rm rel},\,\eta^{\rm rel},\,\phi^{\rm rel})$ and the corresponding absolute kinematics $(p_\mathrm{T},\,\eta,\,\phi)$. All the distributions are normalized to unit area. The lower panels show the ratio of generated-to-validation distributions with statistical uncertainties. The gray bands delimit $\pm5\%$ and $\pm10\%$ around unity.}
    \label{fig:figOO_particle1D}
\end{figure*}

\subsection{Event- and particle-level validation} 

Figures 2 and 3 summarize the \textbf{one-dimensional performance} for the Stage-1 generator at both event- and particle-level. 

At the \textbf{event-level (Fig. 2)}, we validate the one-dimensional distributions of the global properties used for conditioning and generation. The displayed variables include the event-level mean transverse-momentum $\langle p_\mathrm{T}\rangle$ and pseudorapidity $\langle \eta \rangle$, the event-plane angle $\psi_2$, the impact parameter $b$ and multiplicity component $N_{\rm particles}$. 
The generated distributions closely follow the validation reference across the bulk of the phase space, establishing that the model is successfully reproducing the global event characteristics. 

At the \textbf{particle level (Fig. 3)}, we validate the one-dimensional distributions of the six particles representations, including both the relative variables and the corresponding absolute kinematics. 
Good agreement is observed simultaneously in both the relative and absolute variables: the model reproduces the central region, and widths of the distributions, with the Gen/Val ratio remaining close to unity over the dominant phase-space region. The performance of the relative variables is particularly important, as it indicates that the generator captures event-normalized structure rather than simply matching global averages. 
Fluctuations are visible in the extreme high-$p_\mathrm{T}$ tail, where statistical uncertainties are large because such particles are rare (only $0.0032\%$ of validation particles satisfy $p_\mathrm{T}>10~\mathrm{GeV}/c$). The impact of these tails is assessed in Section 4.4 through the end-to-end jet reconstruction and substructure observables. 

We perform \textbf{two-dimensional consistency checks} to verify that the Stage-1 generator successfully reproduces the joint structure of the most important event-level and particle-level observables. Figure 4 shows the generated-to-validation ratio maps, with the 95th-percentile region of the validation distribution delimited by the dashed lines for the event-level and the particle-level features respectively. 
These validations are designed to probe whether the generator preserves correlations between global activity variables and event-shape proxies, not only their one-dimensional projections. In the populated phase-space region, the ratio is compatible with unity within statistical precision, while larger fluctuations appear primarily in low-occupancy edge bins, as expected from limited statistics in the tails. 

To summarize the agreement quantitatively, we compute a binned generated-to-validation score in the full region and report both the mean and the median over bins (the median being more robust to sparsely populated edge bins). For the event-level pairs, we obtain $(\overline{r},\,\tilde r)=(1.02,\,1.00)$ for $\langle\eta\rangle$ vs $\langle p_\mathrm{T}\rangle$ and $(0.97,\,0.99)$ for $N_{\rm particles}$ vs $\sum p_\mathrm{T}$. At the particle-level, the corresponding scores are $(1.01,\,1.00)$ for both $\eta$ vs $p_\mathrm{T}$ and for $\phi$ vs $p_\mathrm{T}$, consistent with the near-unity ratio maps in the densely populated region. 
This confirms that the Stage-1 model reproduces not only the inclusive distributions but also the dominant inter-variable structure that controls event classification and mixing-pool definitions in mixed-event workflows. 



\subsection{Event-particle correlations and collective structure}

One of the key requirements for validating the generated events is that particle-level kinematics remain consistent with event-level. In particular, the generator must preserve the correlation between the stored event features and the azimuthal organization of the generated particle ensemble. We therefore perform a set of event-particle correlation tests that explicitly connect particle-level observables to event-level quantities beyond one-dimensional closure. 

\paragraph{Q-vector evaluation for particle- to event-level correlation} 

We first test whether the event-level second-harmonic flow vector $(Q_x,Q_y)$ produced by the event branch is consistent with the same quantity reconstructed from the generated particles. For each generated event we recompute $Q_x$, $Q_y$, using the particle-level $(p_\mathrm{T},\phi)$ features. 
The left panel of Figure 5 shows the reconstructed $(Q_x^{\rm reco},Q_y^{\rm reco})$ from particles as a function of the generated event-level $(Q_x^{\rm gen},Q_y^{\rm gen})$. 
A strong linear correspondence is observed between the reconstructed and generated $Q$, with correlation coefficients $\rho\simeq 0.99$ for both components. The $\Delta Q_{x,y}=Q_{x,y}^{\rm reco}-Q_{x,y}^{\rm gen}$ are shown on the right panel of Figure 5, where the distribution is narrow and centered close to zero, with $(\mu_x,\sigma_x)=(0.09,\,1.16)$ and $(\mu_y,\sigma_y)=(-0.03,\,1.27)$. This indicates that the generated particle ensemble reflects the global second-harmonic azimuthal structure encoded by the event branch, thereby preserving the overall event topology on the generated particle level. 

\begin{figure*}[!htb]
    \centering
    \includegraphics[width=0.9\linewidth]{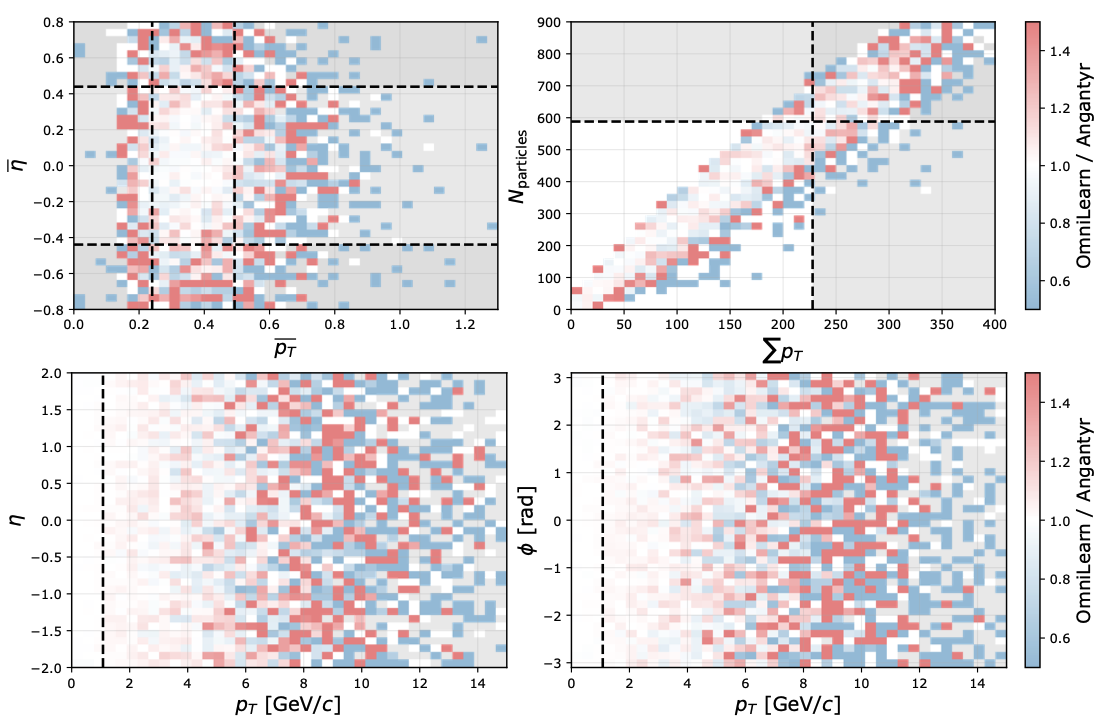}
    \caption{Two-dimensional generated-to-validation ratio maps probing joint structures at both \textbf{(top)} event- and \textbf{(bottom)} particle-level in O-O collisions. The dashed lines indicate the interval containing 95\% of the statistics, and are shown as a visual guide to highlight the region where most of the data populate and where the agreement between generated and reference distributions is most relevant.} 
\end{figure*}

\begin{figure*}[!htb] 
    \centering
    \includegraphics[width=0.75\linewidth]{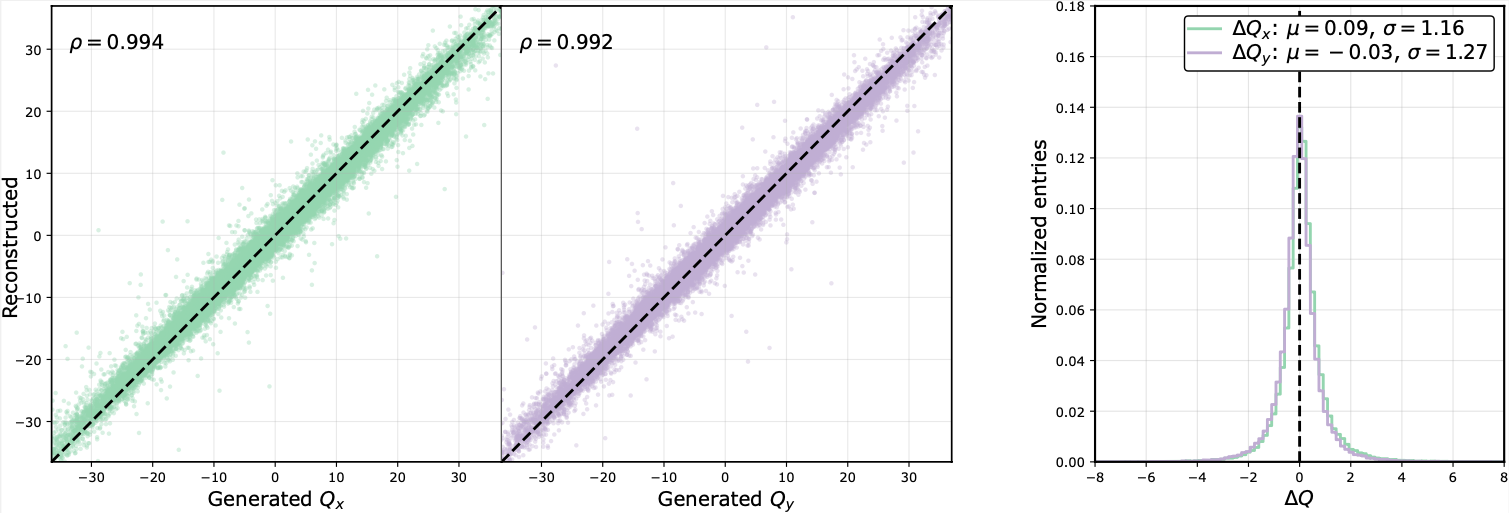}
    \caption{Event-particle consistency checks for the Stage-1 O-O generator. \textbf{Left:} correlation between event-level $Q$-vector components $(Q_x,Q_y)$ generated by the event branch and the corresponding values reconstructed from the generated particle ensemble. \textbf{Right:} the corresponding $\Delta Q_{x,y} = Q_{x,y}^{\mathrm{reco}}-Q_{x,y}^{\mathrm{gen}}$ distributions.}
    \label{fig:figOO_merged}
\end{figure*}

\paragraph{Two-particle azimuthal correlations}

Next, we probe whether the generator reproduces the azimuthal correlation structure of particle pairs in a trigger-selected configuration. We study the distribution of $\Delta\phi=\phi_{\rm trig}-\phi_{\rm assoc}$ using trigger with $p_\mathrm{T}^{\rm trig}>4~\mathrm{GeV}/c$ and all associated particles within the same event. The left panel of Figure 6 shows the $\Delta\phi$ distribution. 
Good agreement between validation and generated samples is maintained across the full $\Delta\phi\in[-\pi,\pi]$ interval, with the generated-to-validation ratio consistent with unity at the $\sim$2\% level in the populated region. To further test the conditional structure, we examine the joint distribution of $\Delta\phi$ versus trigger $p_\mathrm{T}$ (Figure 6, right). In the high-occupancy region, the ratio map stays consistent with unity, showing that the model preserves the $\Delta\phi$ correlation and its dependence on the trigger transverse momentum $p_\mathrm{T}^{\rm trig}$. This is important for correlation measurements and for mixed-event pool definitions that incorporate event activity and trigger selections. 

\paragraph{Collectivity study with $v_2\{SP\}(p_\mathrm{T})$}
A key requirement for a heavy-ion event generator is to reproduce not only single-particle kinematics, but also the long-range collective anisotropies that encode the medium response to the initial-state geometry. The elliptic-flow coefficient $v_2(p_\mathrm{T})$ is the dominant harmonic in non-central collisions and a standard benchmark for collectivity, as it reflects the correlation of particle emission with a common symmetry plane and its event-by-event fluctuations. We therefore validate the generator using a two-subevent scalar-product (SP) measurement \cite{sp_method} with an $\eta$ gap, which suppresses short-range non-flow correlations by correlating particles with a reference estimated in a separated pseudorapidity region. Each event is split into two sub-events $A$ and $B$ in pseudorapidity, and for each sub-event a second-harmonic reference is constructed from the azimuthal distribution of particles using $p_\mathrm{T}$ weights. In a SP formulation this corresponds to correlating the unit flow vector of the particle of interest with the subevent reference and normalizing by the correlation between the two subevent references. For particles in a given $p_\mathrm{T}$ bin, the SP estimator is: 
\begin{align*}
v_2\{SP\}(p_\mathrm{T}) \equiv \frac{ \left\langle \cos\!\left[2\left(\phi - \psi_{2,\mathrm{ref}}\right)\right]\right\rangle_{p_\mathrm{T}} }{\sqrt{\left\langle \cos\!\left[2\left(\psi_{2,A}-\psi_{2,B}\right)\right]\right\rangle }},
\end{align*}
where $\psi_{2,\mathrm{ref}}=\psi_{2,B}$ for particles of interest in sub-event A and $\psi_{2,\mathrm{ref}}=\psi_{2,A}$ for particles of interest in sub-event B. Figure 7 shows $v_2\{SP\}(p_\mathrm{T})$ for both the validation and the generated samples. 
The generator reproduces the overall magnitude and $p_\mathrm{T}$ dependence within uncertainties over the bulk region. This confirms that the model preserves the event-by-event coupling between a global elliptic anisotropy reference and the differential particle emission pattern in a way that survives an $\eta$-separated scalar-product definition.

\begin{figure*}[!htb] 
    \centering
    \includegraphics[width=0.8\linewidth]{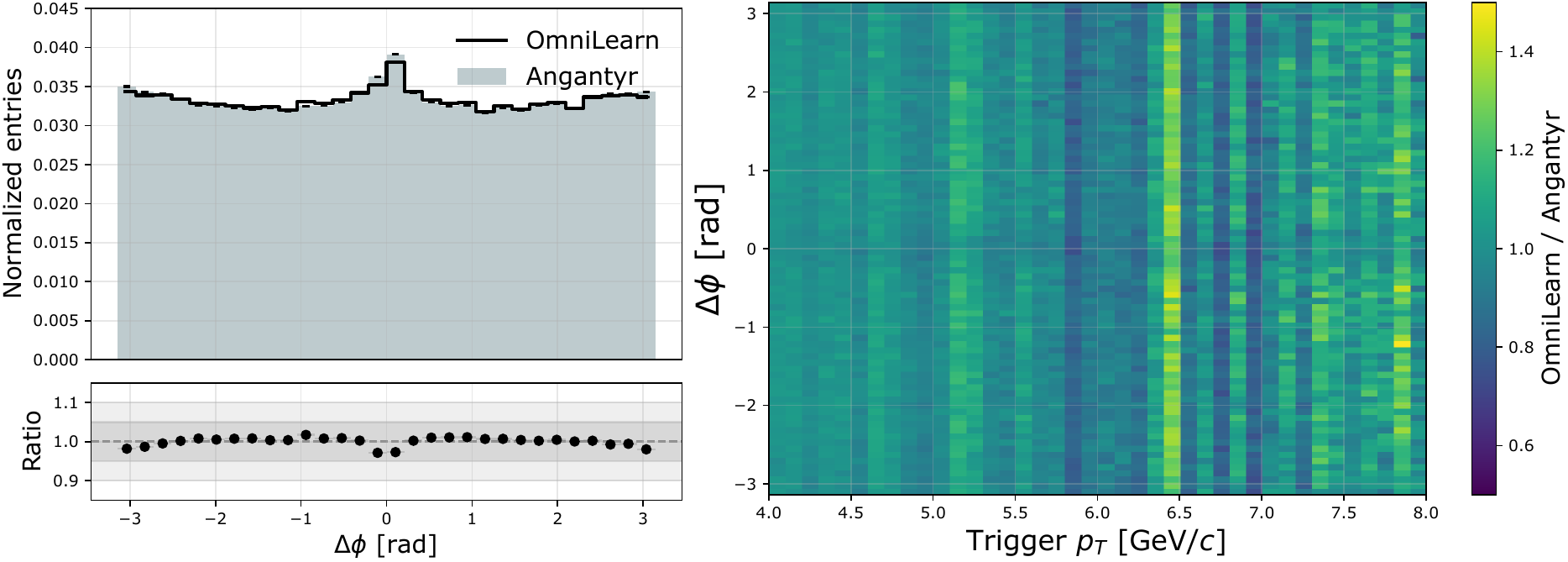}
    \caption{Two-particle azimuthal correlations in O-O collisions.
\textbf{Left:} one-dimensional $\Delta\phi=\phi_{\rm trig}-\phi_{\rm assoc}$ distribution for triggers with $p_\mathrm{T}^{\rm trig}>4~\mathrm{GeV}/c$ and all associated particles, comparing validation (gray) with the generated (black) data. The lower panel shows the generated-to-validation ratio with statistical uncertainties. The gray bands delimit $\pm5\%$ and $\pm10\%$ around unity.
\textbf{Right:} the generated-to-validation ratio map of the distribution $(\Delta\phi, p_\mathrm{T}^{\rm trig})$. The ratio remains consistent to unity within statistical uncertainties.} 
    \label{fig:figOO_dphi}
\end{figure*}

\begin{figure}[!htb] 
    \centering
    \includegraphics[width=0.9\linewidth]{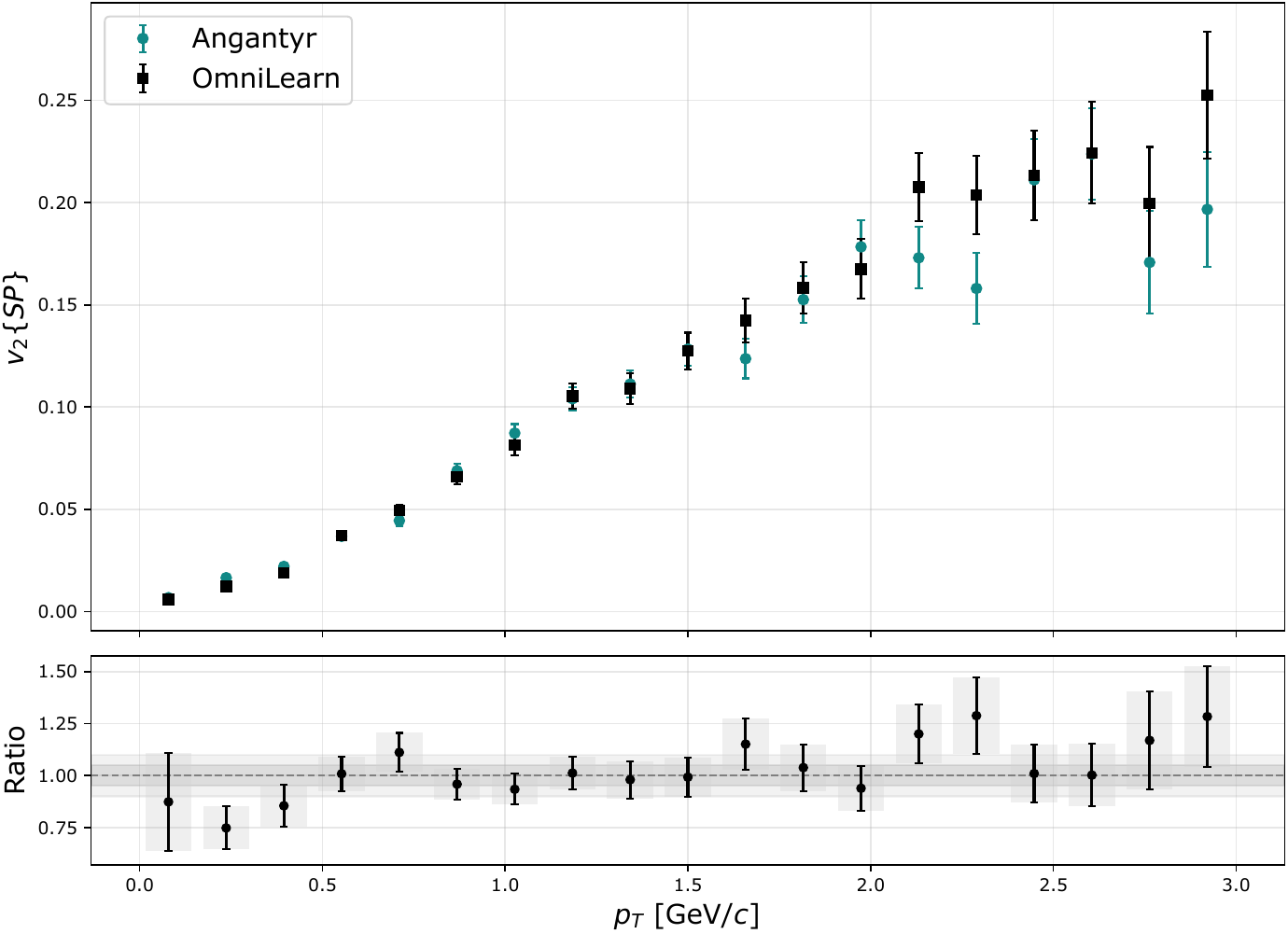}
    \caption{Two-sub-event scalar-product measurement of elliptic anisotropy in O-O collisions $v_2\{SP\}$ as a function of $p_\mathrm{T}$ for the validation (green) and generated (black) data comparison. The lower panel shows the generated-to-validation ratio with statistical uncertainties. The gray bands delimit $\pm5\%$ and $\pm10\%$ around unity.}
    \label{fig:figOO_v2_pT}
\end{figure}

\subsection{Downstream reconstruction: jets and substructure}

\paragraph{Inclusive jets}
A central requirement for downstream applications is that generated particles remain usable after standard reconstruction steps. We therefore perform an end-to-end closure test in which the generated particles are clustered with the same jet definition as the validation sample, using the anti-$k_T$ algorithm \cite{antikt} with radius parameter $R=0.4$ implemented in FastJet \cite{fastjet}. 
Figure 8 summarizes the result of the jet clustering with one-dimensional validation at the reconstructed-jet level for inclusive jet kinematics and basic structural properties: $p_\mathrm{T}^{\rm jet}$, $\eta^{\rm jet}$, $\phi^{\rm jet}$, $E^{\rm jet}$, jet mass, and the number of constituents. Across the bulk of the distributions, the generator reproduces the reconstructed jet spectra and shapes with the generated-to-validation ratios close to unity, indicating that the particle-level closure established in Section 4.2 survives clustering and is not spoiled by reconstruction nonlinearities. 

The agreement is particularly strong for the geometric jet coordinates: $\eta^{\rm jet}$ and $\phi^{\rm jet}$ remain flat and consistent with the validation reference within uncertainties across the acceptance. For $E^{\rm jet}$, jet mass, and constituent multiplicity, the ratios remain compatible with unity over the dominant support of the distributions, demonstrating that the model captures not only the inclusive jet yield but also the internal energy sharing and multiplicity patterns that control basic substructure sensitivity.

In the extreme high-$p_\mathrm{T}$ tail of the jet spectrum deviations are most visible where the generated distribution tends to undershoot the validation reference. The agreement remains within the $\pm 10\%$ band over essentially the entire populated region: the generated-to-validation ratio stays close to unity up to approximately the $99.8$ \% of the validation jet-$p_\mathrm{T}$ distribution. The observed undershoot is confined to the final $\sim$0.2\% of jets in the far tail, where statistical uncertainties are intrinsically large and small absolute differences translate into relative fluctuations.

In conclusion, this end-to-end reconstruction is particularly important as it validates the generator \emph{after} applying a standard physics analysis pipeline: jets are reconstructed from the generated point clouds particle tokens, using the same FastJet anti-$k_T$ algorithm applied to the validation sample, and the resulting observables are directly comparable to those used in experimental analyses. The good agreement across inclusive jet kinematics therefore indicates that the generated particle ensembles are not only distributionally accurate, but also yield consistent physics objects under downstream reconstruction.

\begin{figure*}[!ht] 
    \centering
    \includegraphics[width=0.75\linewidth]{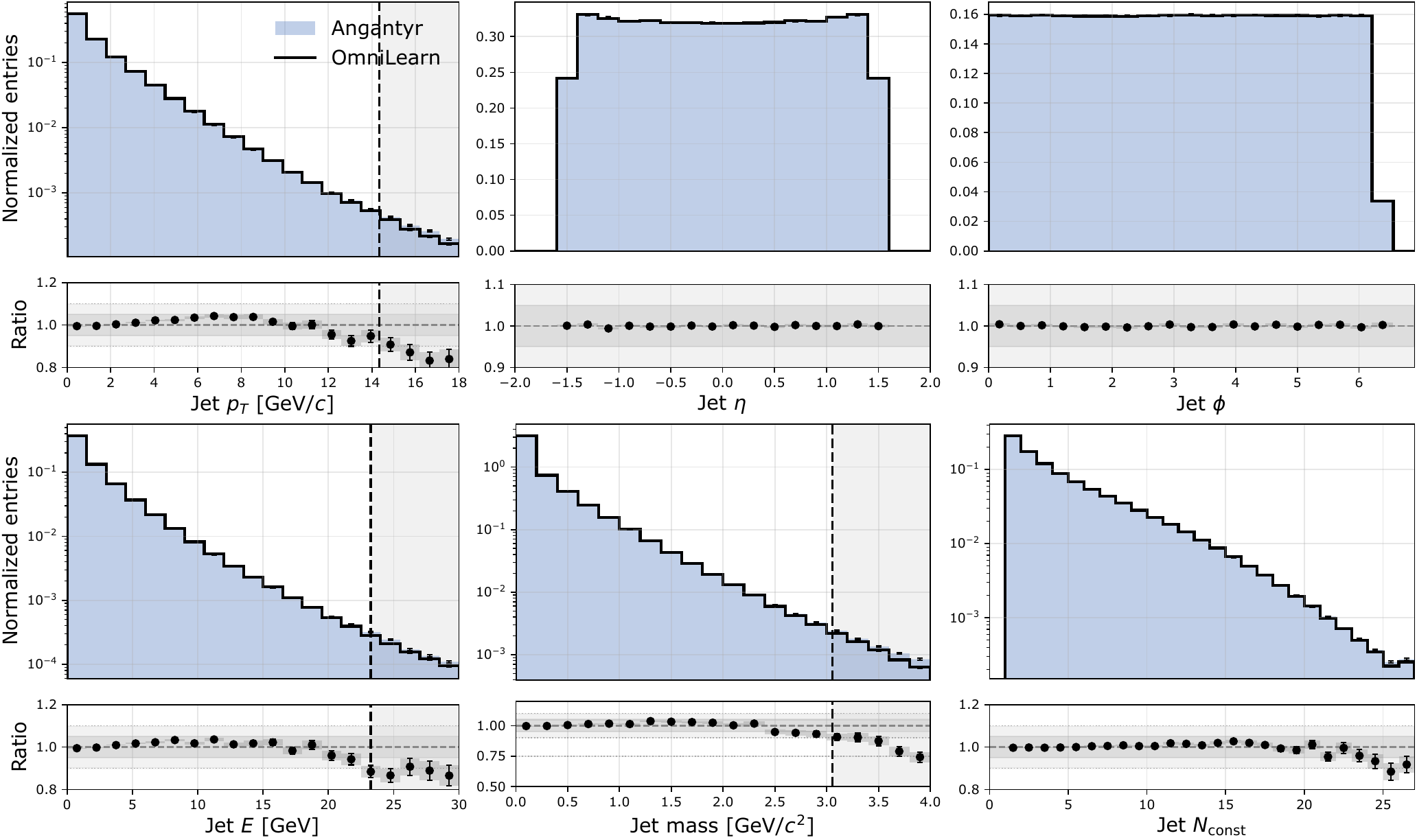} 
    \caption{Jet reconstruction closure in O-O collisions using FastJet anti-$k_T$ jets with $R=0.4$. The top panels compare the validation and generated distributions for $p_\mathrm{T}^{\rm jet}$, $\eta^{\rm jet}$, $\phi^{\rm jet}$, $E^{\rm jet}$, jet mass, and the jet constituent multiplicity. The bottom panels show the generated-to-validation ratios with statistical uncertainties and shaded $\pm5\%$ and $\pm10\%$ bands around unity. The dashed lines delimit the region in which lies 99.8 \% of the statistics.}
    \label{fig:figOO_jets_v1_1D}
\end{figure*}

\paragraph{Jet substructure: recoil-free axis, grooming, and angularities} To tighten the jet-level validation beyond inclusive kinematics, we compare substructure observables that probe correlated radiation inside the jet and can be more sensitive to mis-modelling. Figure 9 shows validation for three representative observables, with the cut $p_\mathrm{T}^{\rm jet}>10~\mathrm{GeV}/c$ applied. (i) A recoil-free axis benchmark using Winner-Take-All (WTA) \cite{wta} re-clustering with Cambridge/Aachen \cite{ca_algo}, quantified by the axis displacement $\Delta R_{\rm WTA-std}\equiv \Delta R(J_{\rm WTA},J)$. (ii) Soft Drop grooming with $z_{\rm cut}=0.1$ \cite{softdrop}, reported as the groomed-to-ungroomed axis displacement $\Delta R_{\rm SD}\equiv \Delta R(J_{\rm SD},J)$. (iii) The jet angularity \cite{angularities} $\lambda_{\alpha}^{\kappa=1}$ with $\alpha=2$,
$\lambda_{\alpha}^{\kappa=1}=\sum_{i\in{\rm jet}} z_i\left(\frac{\Delta R_i}{R}\right)^{\alpha}, \qquad z_i=\frac{p_{\mathrm{T},i}}{p_\mathrm{T}^{\rm jet}}.$ Across the dominant support in Fig. 9, the generated sample reproduces the validation reference with the generated-to-validation ratios consistent with unity, indicating that the model captures not only inclusive jet production but also key features of the internal angular and energy-sharing structure.

\begin{figure*}[!htb] 
    \centering
    \includegraphics[width=0.8\linewidth]{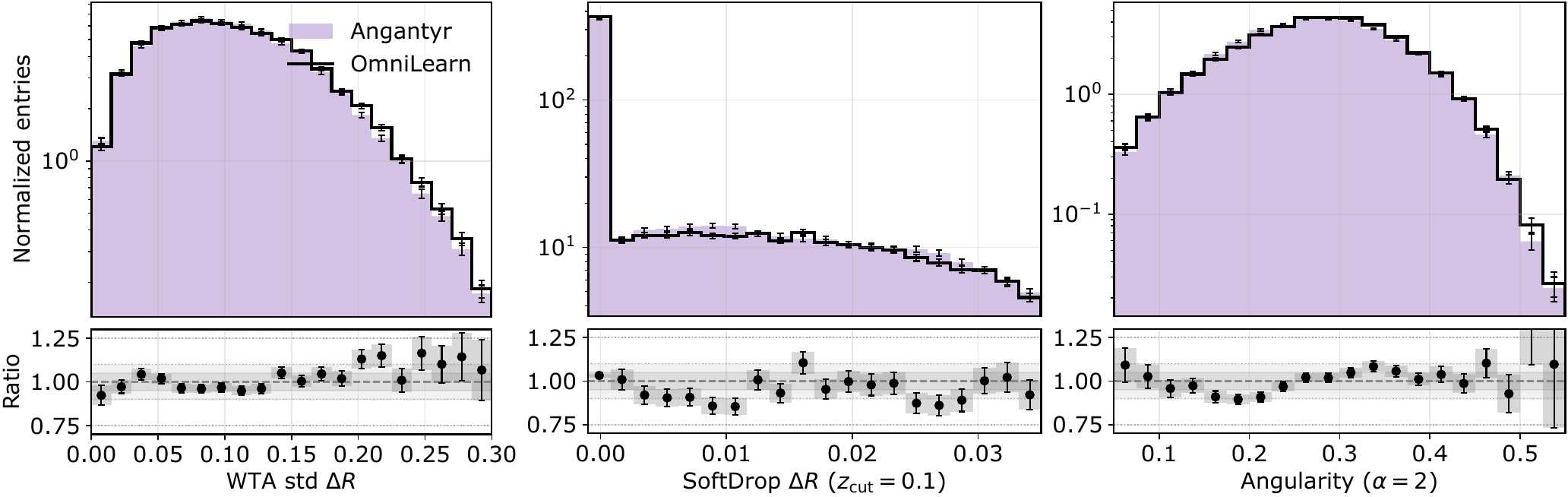} 
    \caption{Distributions for \textbf{(left)} the WTA-standard axis displacement $\Delta R$, \textbf{(middle)} the Soft Drop groomed-to-ungroomed axis displacement $\Delta R$ with $z_{\rm cut}=0.1$ and \textbf{(right)} jet angularity with $\alpha=2$. Lower panels show the generated-to-validation ratios with statistical uncertainties and shaded $\pm5\%$ and $\pm10\%$ bands around unity.}
    \label{fig:figOO_subs_jets}
\end{figure*}

\paragraph{Underlying-event background proxy}
As a final jet-level closure test, we validate the underlying-event (UE) background estimate obtained with the area-median method. We estimate the event-wise background density $\rho$ using a median-based procedure and compute the jet background proxy $\rho A_{\rm jet}$ for reconstructed anti-$k_T$ jets with $R=0.4$ and $p_\mathrm{T}^{\rm jet}>10~\mathrm{GeV}/c$, where $A_{\rm jet}$ is the active jet area. Figure 10 shows the resulting $\rho A_{\rm jet}$ distributions for the validation and generated samples. The generated distribution follows the validation reference over the bulk of the spectrum, with the generated-to-validation ratio remaining close to unity and within the $\pm 10\%$ band across the populated region. This agreement indicates that the generated events reproduce consistent UE background levels and fluctuations when processed through the same median-based estimator and jet-area correction as the validation reference sample. 

\begin{figure}[!htb] 
    \centering
    \includegraphics[width=0.9\linewidth]{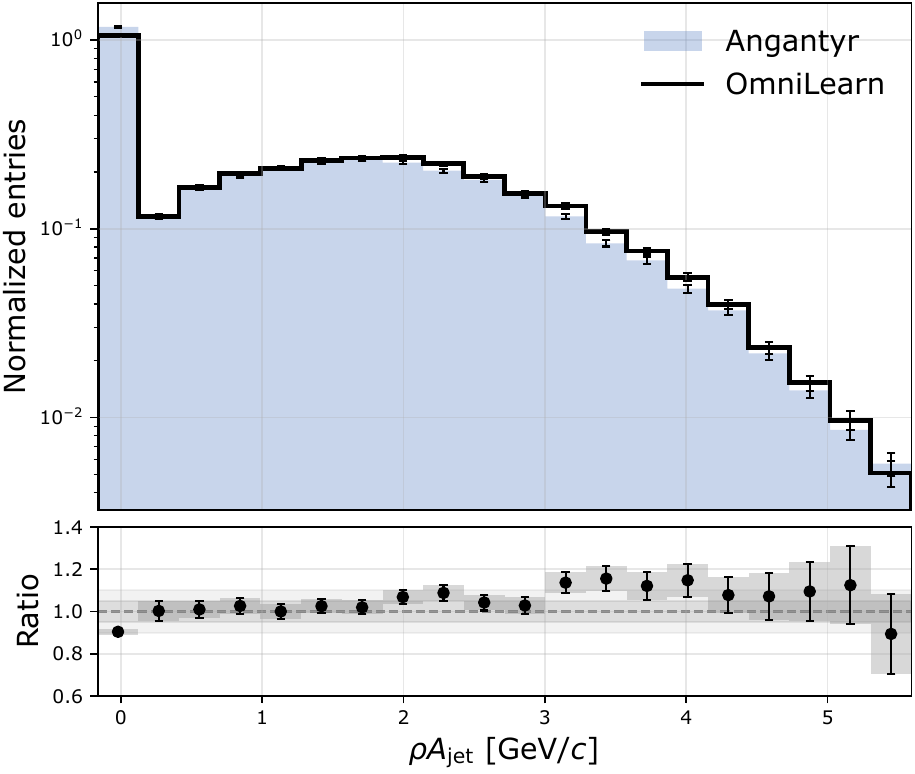} 
    \caption{\textbf{Top:} Event-wise background density $\rho$ is estimated with a median-based $k_T$ procedure (excluding the two hardest jets) and multiplied by the reconstructed jet area $A_{\rm jet}$ for anti-$k_T$ jets with $R=0.4$ and $p_\mathrm{T}^{\rm jet}>10~\mathrm{GeV}/c$. \textbf{Bottom:} generated-to-validation ratios with statistical uncertainties and shaded $\pm5\%$ and $\pm10\%$ bands around unity.}
    \label{fig:figOO_rhoAjet}
\end{figure}

\subsection{Global fidelity metrics}

To complement the observable-by-observable validations presented in the section above, we summarize model quality using compact metrics designed to compare training configurations. 
In particular, we focus here on three models that use the same architecture and pre-processing but differ in available statistics or training time: a model trained on 20\% of the used full dataset, a 1M-training events (full dataset) model trained for 30 epochs (early-stopped), and the final 1M-training events model trained for 70 epochs. 
Figure 11 shows four global indicators: (a) marginal fidelity at the event level, (b-c) a condensed measure of event-particle coherence based on Q-vector closure discussed in Section 4.3, and (d) a particle-level marginal fidelity score. 
For event-level marginal fidelity we report the mean one-dimensional Wasserstein distance averaged over the event observables used as conditioning inputs in this work,
\begin{equation*}
\langle W_1\rangle_{\rm evt} \equiv \frac{1}{N_{\rm var}}\sum_{v} W_1\!\left(p_{\rm val}(v),\,p_{\rm gen}(v)\right),
\end{equation*}
where $v$ runs over the set of event-level variables and $N_{\rm var}$ is the number of variables included in the average. $W_1\!\left(p_{\rm val}(v),p_{\rm gen}(v)\right)$ denotes the one-dimensional Wasserstein distance between the validation and generated distributions of the scalar observable $v$. Discrete count-like variables ($N_{\rm coll}$, $N_{\rm part}$, $N_{\rm particles}$) are rounded to the nearest integer prior to computing distances. 
As displayed in Fig. 11(a), the event-level score improves systematically with increased training statistics and longer training. 

To summarize conditional consistency between the generated particle set and the event-level representation, we use the Q-vector closure discussed in Section 4.3 applied to the different models. We reconstruct the second-harmonic Q-vector from generated particles and compare it to the corresponding event-level values stored in the generated event representation. We compress this comparison into the Pearson correlation ${\rm corr}(Q^{\rm true},Q^{\rm reco})$ (Fig. 11(b)) and the residual scale
${\rm RMS}(Q)\equiv \sqrt{\left\langle\left(Q^{\rm reco}-Q^{\rm true}\right)^2\right\rangle},$
shown in Fig. 11(c), quantifying the typical event-by-event mismatch between the event-level $Q$ and the reconstructed particle-level. 
In both panels we report the average of the $Q_x$ and $Q_y$ scores, with the error bars indicating the spread between the two components. This compact view is useful for distinguishing models that achieve improved event-level marginals but differ in their ability to transmit the conditioning information to particle-level correlations. 
The three models separate clearly in these metrics. The final 1M/70-epoch model achieves near-linear correlations as discussed previously, and RMS errors of order unity (RMS$_{Q_x}=1.14$, RMS$_{Q_y}=0.99$). In contrast, the 1M/30-epoch model shows a pronounced degradation in conditional structure, with correlations dropping to ${\rm corr}(Q_x)=0.888$ and ${\rm corr}(Q_y)=0.901$ and RMS increasing to $\sim 4.4$-$4.6$. The 20\% model is intermediate: correlations remain relatively high (${ \rm corr}(Q_x)=0.952$, ${\rm corr}(Q_y)=0.978$), but the residual scale is still larger than for the final model (RMS$_{Q_x}=3.16$, RMS$_{Q_y}=2.07$).

Finally, to include an explicit particle-level marginal score beyond the Q-closure test, we compute a particle-level mean one-dimensional Wasserstein distance using the single-particle kinematic distributions of $(p_\mathrm{T},\eta,\phi)$: 
\begin{equation*}
\begin{aligned}
\langle W_1\rangle_{\rm part}
&\equiv \frac{1}{3}\Big[
W_1\!\left(p_\mathrm{T}^{\rm val},p_\mathrm{T}^{\rm gen}\right)
+W_1\!\left(\eta^{\rm val},\eta^{\rm gen}\right) \\
&\qquad\quad
+W_1\!\left(\phi^{\rm val},\phi^{\rm gen}\right)
\Big],
\end{aligned}
\end{equation*}
where particle samples are assembled over events after masking padded entries. For the azimuthal angle, periodicity is handled by computing the distance in the $(\sin\phi,\cos\phi)$ representation. 
For the three models, we find $\langle W_1\rangle_{\rm part}\simeq 2.25\times 10^{-3}$ (20\%), $2.86\times 10^{-3}$ (1M/30 epochs), and $2.18\times 10^{-3}$ (1M/70 epochs), as displayed in Figure 11(d). The particle-marginal score varies only weakly across configurations, indicating that single-particle kinematics are captured relatively early, whereas the dominant gains from increased statistics and longer training are reflected in improved event-level marginals and, more strongly, in event-particle correlation.

Taken together, Figure 11 shows that longer training and larger statistics primarily improve the conditional structure of the generator: the 1M/70-epoch model achieves the best event-level marginal fidelity and, more importantly, the strongest event-particle coherence (highest Q correlation and smallest RMS). In contrast, the particle-level marginal score varies only weakly across the three trainings, indicating that particle-level marginals saturate earlier, whereas the dominant gains from extended training are expressed in event-level agreement and in the faithful propagation of event-level conditioning information to particle-level.
\begin{figure*}[!htb] 
    \centering
    \includegraphics[width=1.\linewidth]{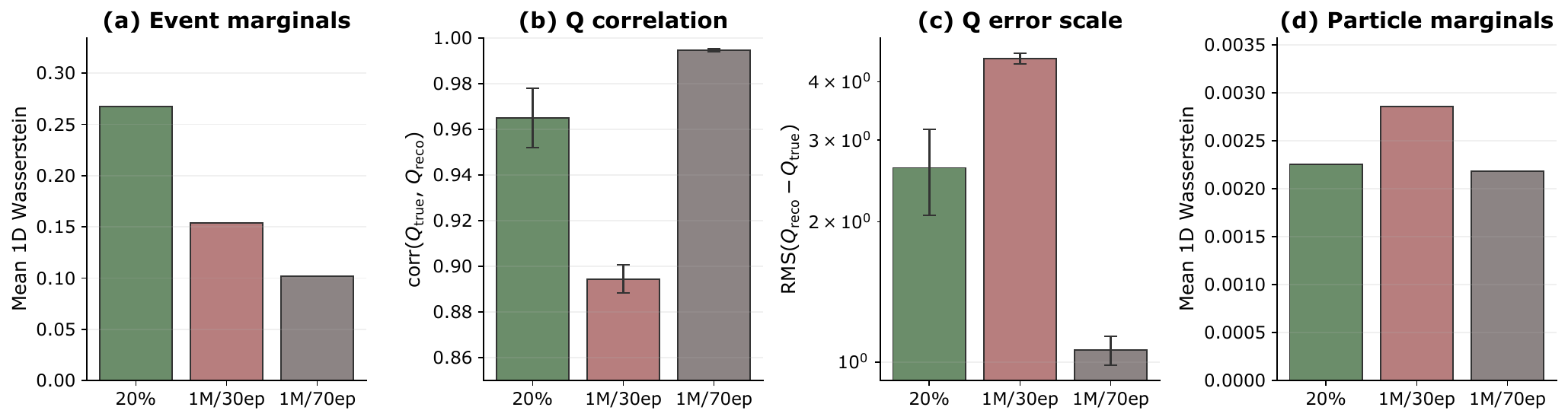} 
    \caption{Global fidelity metrics for O-O models, comparing 20\% statistics, 1M/30 epochs, and 1M/70 epochs. (a) Event-level: mean 1D Wasserstein distance over event conditioning observables. (b-c) Event-particle: Q-vector closure (correlation-RMS). (d) Particle-level: marginal fidelity from the mean 1D Wasserstein distance over $(p_\mathrm{T},\eta,\phi)$.}
    \label{fig:figOO_scores}
\end{figure*}

\paragraph{Classifier-based discriminability test}
As an additional, fully multivariate validation, we train a lightweight supervised classifier to distinguish generated from validation events and particles. 
The classifier performance is then summarized by the receiver operating characteristic (ROC) curve and its area under the curve (AUC), where ${\rm AUC}=0.5$ indicates no separability (random guessing) and larger AUC values quantify increasingly detectable differences between generated and validation samples. 
For the \textbf{event-level test}, we train a multilayer perceptron (MLP) on the event feature vector to classify validation events versus generated events. We use a 70/15/15 train/test/validation split. Input features are normalized using the mean and standard deviation computed on the training subset, and the same transformation is applied to the validation and test subsets.
For the \textbf{particle-level test}, we use an order-invariant classifier suitable for variable-length particle lists. Each particle is processed by the same small MLP to produce an embedding, and these embeddings are averaged over particles using a mask to ignore padding, similar to the \textsc{Deep Sets}~\cite{Zaheer:2017} architecture. The resulting event embedding is then passed to a final MLP head to classify validation versus generated events. This construction ensures that the classifier output is insensitive to the arbitrary ordering of particles in the input. 

Figure 12 shows that, for the final O-O model, both the event-level and particle-level classifiers yield ROC curves consistent with random guessing (AUC $\simeq 0.5$), indicating that the generated samples are difficult to distinguish from validation in these global representations. 

\begin{figure*}[!htb] 
    \centering
    \includegraphics[width=0.8\linewidth]{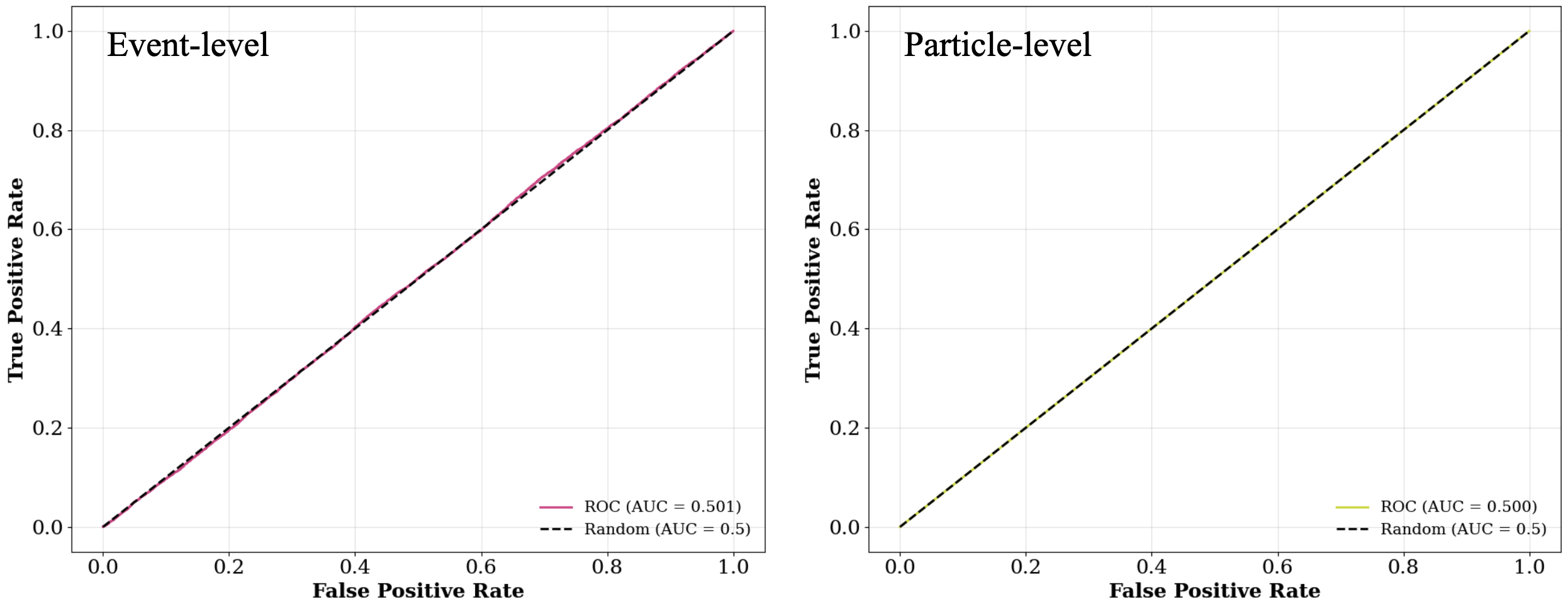} 
    \caption{Classifier-based discriminability test (ROC) between validation and generated samples for the final O-O model. \textbf{Left:} event-level MLP. \textbf{Right:} particle-level order-invariant classifier.}
    \label{fig:figOO_class}
\end{figure*}

\section{Fine tuning with Pb-Pb collisions}

We now extend to the two-stage training strategy by fine-tuning the Stage-1 O-O generator on Pb-Pb collisions. This step targets the high-multiplicity regime, where preserving global coherence between the event-level conditioning variables and the particle ensemble becomes substantially more challenging. The Pb-Pb training uses the same OmniLearn architecture and pre-processing pipeline as described for O-O collisions in Section 4, and the same HDF5 data format. The used Pb-Pb sample corresponds to 10\% of the O-O statistics. 
Fine-tuning is performed on the Perlmutter supercomputer in data-parallel mode using Horovod across 5 GPU nodes (20 GPUs total), same as the Stage-1 training. We initialize the model from the best O-O checkpoint and train for 25 epochs, saving the best-performing weights according to the validation loss. Due to the substantially larger per-event multiplicities in Pb-Pb, GPU memory constraints require a per-rank batch size of 1.

\subsection{Validations at event- and particle-level}

We first assess whether the fine-tuned model reproduces the marginal distributions of key Pb-Pb collisions observables. Figure 13 shows event-level validations for the following representative global quantities: the impact parameter $b$, the number of binary nucleon-nucleon collisions $N_{\rm coll}$, and the multiplicity $N_{\rm particles}$. 
As expected for Pb-Pb collisions, these observables span a broad dynamical range and are dominated by large event-by-event fluctuations associated with varying collision geometry. The generated sample reproduces the dominant support of all three distributions, with the generated-to-validation ratios remaining compatible with unity over the phase space. 
More precisely, the Pb-Pb fine-tuning extends the accessible geometry and activity ranges by more than an order of magnitude relative to the O-O model: the impact parameter distribution reaches $b\sim 18$ (compared to $b\lesssim 10$ in O-O), $N_{\rm coll}$ extends up to $\sim 3\times10^{3}$ (compared to $\sim 10^{2}$), and the multiplicity spans up to $N_{\rm particles}\sim 10^{4}$ (compared to $\sim 10^{3}$). 

\begin{figure*}[!htb] 
    \centering
    \includegraphics[width=0.9\linewidth]{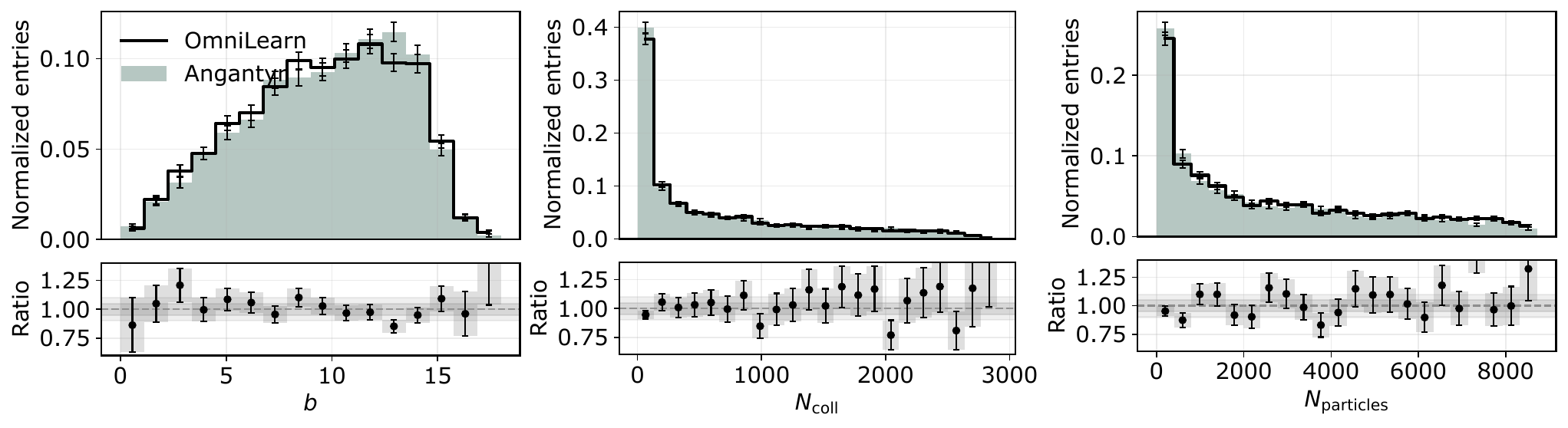} 
    \caption{\textbf{Top:} distributions shown for the impact parameter $b$, the number of binary collisions $N_{\rm coll}$, and the multiplicity $N_{\rm particles}$ for both generated and validation samples. They are normalized to unit area. \textbf{Bottom:} The generated-to-validation ratios with statistical uncertainties and shaded $\pm5\%$ and $\pm10\%$ bands around unity.}
    \label{fig:event1D_}
\end{figure*}

The particle kinematics $(p_\mathrm{T},\eta,\phi)$ are validated in Figure 14 after masking padded entries. The generated sample reproduces the steeply falling Pb-Pb transverse-momentum spectrum over the dominant support, with the generated-to-validation ratio remaining close to unity through the bulk and deviations becoming visible only in the far high-$p_\mathrm{T}$ tail where statistical uncertainties are largest. The pseudorapidity and azimuthal distributions are very well described: within the track acceptance, $\eta$ is reproduced with a flat generated-to-validation ratio at the percent level, and $\phi$ remains uniform and consistent with the validation, indicating that the fine-tuned model does not introduce spurious longitudinal or azimuthal distortions at the particle level. 

\begin{figure*}[!htb] 
    \centering
    \includegraphics[width=0.9\linewidth]{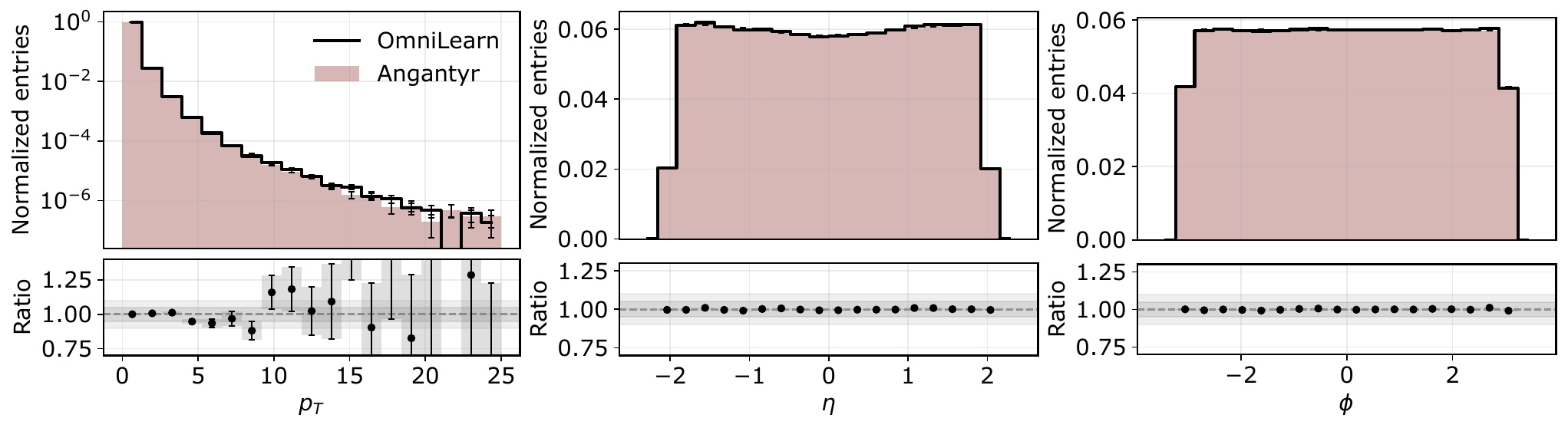} 
    \caption{\textbf{Top:} The absolute single-particle kinematic $(p_\mathrm{T},\eta,\phi)$ distributions for both validation and generation samples. The distributions are normalized to unit area. 
    \textbf{Bottom:} The generated-to-validation ratios with statistical uncertainties and shaded $\pm5\%$ and $\pm10\%$ bands around unity.}
    \label{fig:event1D_}
\end{figure*}

\subsection{Downstream reconstructions}

\paragraph{Inclusive jets}
We next validate that the fine-tuned Pb-Pb generator remains consistent under standard jet finding. Anti-$k_T$ jets with $R=0.4$ are clustered from the generated particle ensembles with the same FastJet configuration as for the validation sample. Figure 15 (left) compares reconstructed jet observables, including $p_\mathrm{T}^{\rm jet}$, $\eta^{\rm jet},\phi^{\rm jet}$, jet energy, jet mass, and constituent multiplicity. The generator captures the dominant support of the inclusive jet kinematics, with the generated-to-validation ratios remaining in agreement with unity over the bulk of each distribution. The geometric jet coordinates show particularly stable closure: $\eta^{\rm jet}$ and $\phi^{\rm jet}$ are flat and consistent with the validation reference, indicating that the fine-tuning does not introduce any acceptance distortions or azimuthal biases at the reconstructed level. Compared to O-O (Section 4.4), the Pb-Pb distributions extend to larger jet activity and significantly higher constituent multiplicities, reflecting the denser underlying environment; the fact that the closure remains stable across this broadened phase space provides a non-trivial end-to-end validation of the Pb-Pb fine-tuning. Deviations become more pronounced only in the most extreme high-tails, which affects around 0.2\% of the data.

\begin{figure*}[!htb]
    \centering
    \begin{minipage}[t]{0.62\linewidth}
        \centering
        \includegraphics[width=\linewidth]{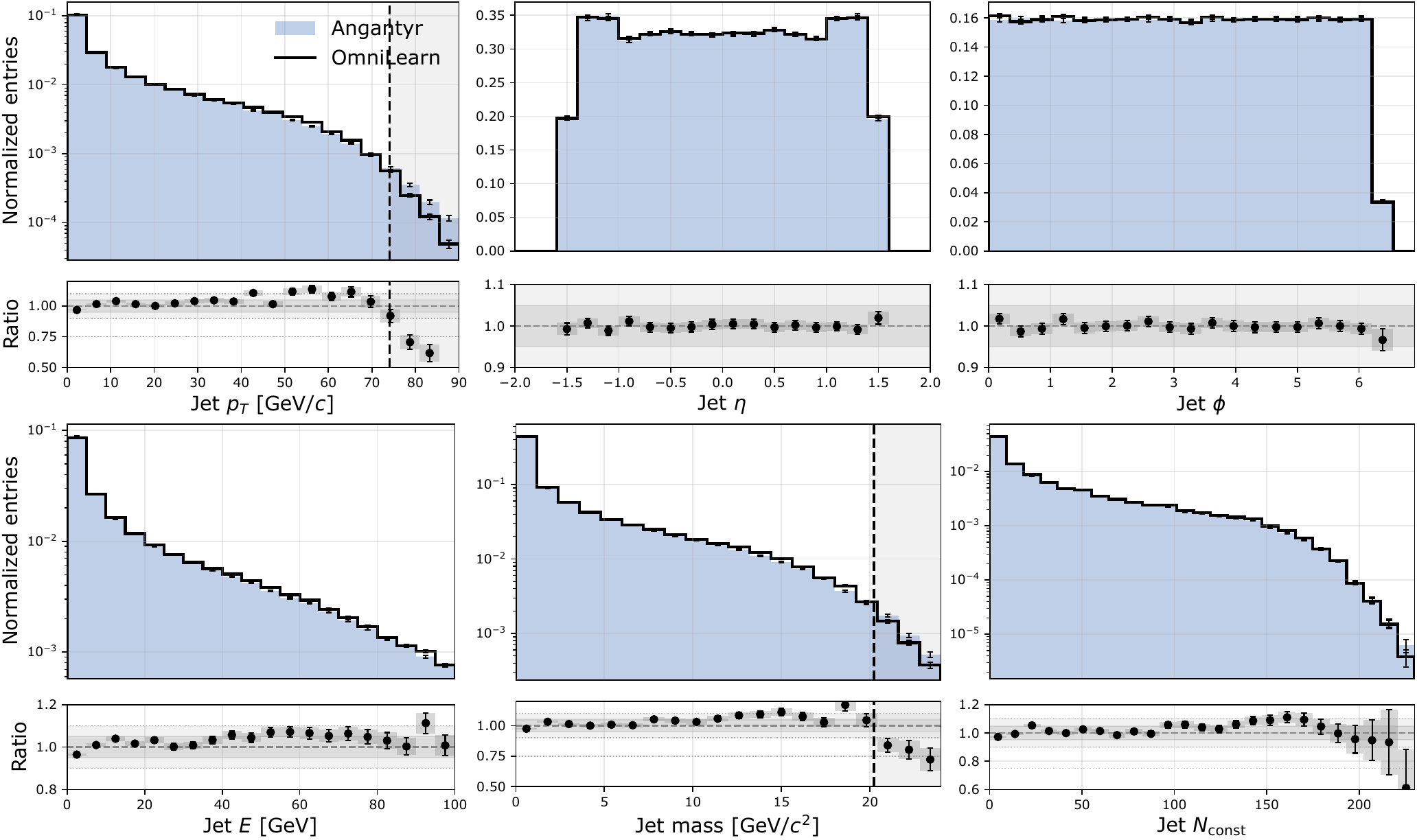}
    \end{minipage}\hfill
    \begin{minipage}[t]{0.34\linewidth}
        \centering
        \includegraphics[width=\linewidth]{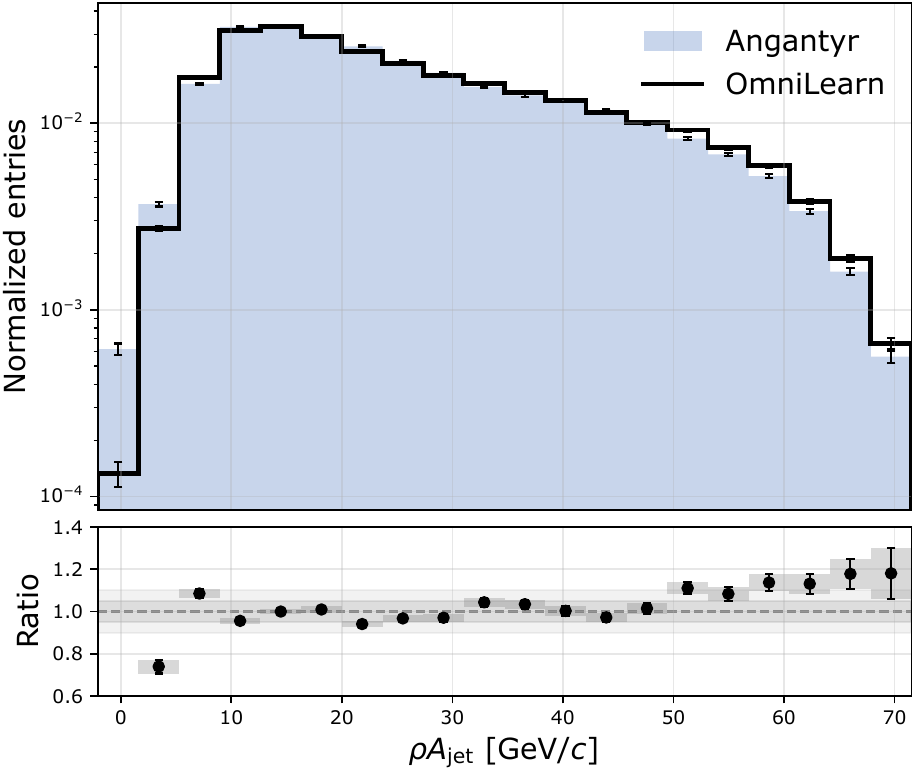}
    \end{minipage}
    
    \caption{\textbf{Left:} Inclusive anti-$k_T$ reconstructed jet observables reconstructed with $R=0.4$ (top), with the generated-to-validation ratios (bottom) displayed for each of the observables with statistical uncertainties. The dashed lines delimit the region in which lies 99.5 \% of the statistics. \textbf{Right:} Jet-level background proxy $\rho A_{\rm jet}$ for $p_\mathrm{T}^{\rm jet}>10~\mathrm{GeV}/c$.}
    \label{fig:figpbpb_jets_rhoA}
\end{figure*}

\paragraph{Underlying-event background proxy}
A distinctive feature of jet reconstruction in high-multiplicity Pb-Pb collisions is the need to account for the underlying-event background. We therefore validate the per-jet background proxy $\rho A_{\rm jet}$, where $\rho$ consists of the event-wise median background density estimator and $A_{\rm jet}$ the active jet area. 
Figure 15 (right) shows the $\rho A_{\rm jet}$ distribution for jets, with the cut $p_\mathrm{T}^{\rm jet}>10~\mathrm{GeV}/c$ applied, for both the validation and generated samples, together with the generated-to-validation ratio. The generator reproduces the shape and normalization over the populated region, with the ratio remaining close to unity within the $\pm10\%$ band for most of the distribution. 
Relative to O-O collisions, the $\rho A_{\rm jet}$ spectrum is substantially broader and extends to larger values, reflecting the stronger underlying-event background in Pb-Pb collisions. The observed closure thus indicates that the fine-tuned model reproduces this new learned background environment consistently when processed through the same median-density and jet-area pipeline as the validation sample.

\subsection{Physics-informed loss}

A central requirement for a realistic heavy-ion event generator is that the produced particles collectively encode the azimuthal anisotropy specified by the event-level conditioning. This collective structure is characterized by the flow vector $Q$ and the associated event-plane angle, as discussed in Section 4.3. Thus, for the model to be physically meaningful, the $Q$-vector components and event-plane angle reconstructed from the generated particles must correlate, event by event, with the generated values. 
In the fine-tuned Pb-Pb model, however, this coherence is initially absent. The $Q$-vector correlations collapse to $\rho_{Q_x} = 0.11$ and $\rho_{Q_y} = 0.06$, and the reconstructed event-plane angle is uniformly distributed relative to the conditioned value: $\langle\cos(2\Delta\psi_2)\rangle = 0.028$. This breakdown originates in the per-particle training objective. The mean-squared-error loss treats each particle independently and imposes no explicit constraint on the collective azimuthal structure of the generated event. In O-O collisions, with multiplicities of $\mathcal{O}(10^3)$ and batch sizes of 64, the gradient signal propagated through the attention mechanism is sufficient for the model to implicitly learn the global $\phi$ modulation on the particle-level. 
However, in Pb-Pb, with multiplicities of $\mathcal{O}(10^4)$ and per-rank batch size of 1 due to memory constraints, each particle's azimuthal coordinate contributes a vanishingly small fraction to the overall flow pattern, and the standard loss carries insufficient information to enforce the collective constraint. 
To address this, we introduce a physics-informed auxiliary loss $\mathcal{L}_\psi$ that directly enforces collective azimuthal alignment. At each training step, the de-noised particle estimate $\hat{x}_0$ obtained from the $v$-prediction (Section 2.2) is reverted to physical $p_\mathrm{T}$ and $\phi$ coordinates, from which we reconstruct the corresponding event-plane angle $\psi_2^\mathrm{reco}$. The loss penalizes the angular misalignment with the conditioned value:
\begin{equation*}
    \mathcal{L}_\psi = 1 - \cos\!\bigl(2(\psi_2^\mathrm{reco} - \psi_2^\mathrm{true})\bigr)\,
\end{equation*}
This loss term is applied only at low noise levels ($t < 0.5$), where the de-noised estimate is physically meaningful. 
Starting from the O-O Stage-1 model, four epochs of training with $\mathcal{L}_\psi$ enabled raise the flow-vector correlations from $(\rho_{Q_x},\,\rho_{Q_y}) \approx (0.11,\,0.06)$ to $(0.74,\,0.82)$, and the angular alignment from $\langle\cos(2\Delta\psi_2)\rangle = 0.04$ to~$0.87$. This demonstrates that a physics-informed constraint targeting the collective flow structure can recover azimuthal coherence that the standard data-driven objective misses at this multiplicity scale. The rapid convergence suggests that the model has already learned the relevant particle-level features during standard fine-tuning; the auxiliary loss serves to align the collective output with the conditioning.

\section{Conclusions}

We have presented a score-based diffusion generative model for the generation of heavy-ion collision events, built on the OmniLearn framework and the Point-Edge Transformer architecture. The model employs a two-stage training strategy: initial training on O-O events, followed by fine-tuning on Pb-Pb events at $\sqrt{s_{\mathrm{NN}}} = 5.36\;\mathrm{TeV}$. 

For O-O collisions, the model achieves excellent performance across a large set of the studied observables. The generated events agree with the training data across event-level properties, particle-level kinematics, two-particle correlations, and reconstructed jet observables including substructures. The model correctly captures the differential elliptic flow $v_2\{\mathrm{SP}\}(p_\mathrm{T})$, and the event-by-event $Q$-vector correlations exceeding 0.99, evidence that the model learns the azimuthal structure underlying flow. Event generation takes around 2.9 s per O-O event on a single NVIDIA A100 GPU, providing a speedup of one to two orders of magnitude compared to transport-model generators \cite{ampt}.

Fine-tuning on Pb-Pb collisions produces similarly good agreement in event- and particle-level distributions, jet properties and their substructure. However, the event-by-event $Q$ reconstruction from generated particles disagrees with expected values.  We attribute this observation to the extreme multiplicity and memory-limited batch size, which limits the gradient propagation below the level needed to enforce the global azimuthal structure. We then introduce a physics-informed auxiliary loss $\mathcal{L}_\psi$ in Section 5.3 that recovers this correlation, increasing $\langle\cos(2\Delta\psi_2)\rangle$ from 0.04 to 0.87 in only four training epochs. This result illustrates how targeted physics constraints can complement data-driven diffusion training in regimes where collective structure is too dilute for the standard training to capture. 

\section*{Code availability}
The source code for this work is available at \cite{code}.

\begin{acknowledgments} 
This research used resources of the National Energy
Research Scientific Computing Center (NERSC), a U.S. Department of Energy Office of Science facility operating under awards NP-ERCAP0033891 and NP-ERCAP0031584. 
MP and RS are supported by the U.S. Department of Energy, Office of Science, Office of Nuclear Physics, under the contract DE-AC02-05CH11231.

\end{acknowledgments}

\bibliographystyle{apsrev4-2}
\bibliography{references}

@techreport{hllhc,
    author = "Zurbano Fernandez, I. and others",
    editor = {B{\'e}jar Alonso, I. and Br{\"u}ning, O. and Fessia, P. and Rossi, L. and Tavian, L. and Zerlauth, M.},
    title = "{High-Luminosity Large Hadron Collider (HL-LHC): Technical design report}",
    reportNumber = "CERN-2020-010",
    doi = "10.23731/CYRM-2020-0010",
    volume = "10/2020",
    month = "12",
    year = "2020"
}

@article{omnilearn,
   title={Method to simultaneously facilitate all jet physics tasks},
   volume={111},
   ISSN={2470-0029},
   url={http://dx.doi.org/10.1103/PhysRevD.111.054015},
   DOI={10.1103/physrevd.111.054015},
   number={5},
   journal={Physical Review D},
   publisher={American Physical Society (APS)},
   author={Mikuni, Vinicius and Nachman, Benjamin},
   year={2025},
   month=mar}

@misc{score_sde,
      title={Score-Based Generative Modeling through Stochastic Differential Equations}, 
      author={Yang Song and Jascha Sohl-Dickstein and Diederik P. Kingma and Abhishek Kumar and Stefano Ermon and Ben Poole},
      year={2021},
      eprint={2011.13456},
      archivePrefix={arXiv},
      primaryClass={cs.LG},
      url={https://arxiv.org/abs/2011.13456}, 
}

@misc{omnilearn_eic,
      title={Point cloud-based diffusion models for the Electron-Ion Collider}, 
      author={Jack Y. Araz and Vinicius Mikuni and Felix Ringer and Nobuo Sato and Fernando Torales Acosta and Richard Whitehill},
      year={2024},
      eprint={2410.22421},
      archivePrefix={arXiv},
      primaryClass={hep-ph},
      url={https://arxiv.org/abs/2410.22421}, 
}

@misc{vparam,
      title={Progressive Distillation for Fast Sampling of Diffusion Models}, 
      author={Tim Salimans and Jonathan Ho},
      year={2022},
      eprint={2202.00512},
      archivePrefix={arXiv},
      primaryClass={cs.LG},
      url={https://arxiv.org/abs/2202.00512}, 
}

@article{angantyr,
   title={The Angantyr model for heavy-ion collisions in Pythia8},
   volume={2018},
   ISSN={1029-8479},
   url={http://dx.doi.org/10.1007/JHEP10(2018)134},
   DOI={10.1007/jhep10(2018)134},
   number={10},
   journal={Journal of High Energy Physics},
   publisher={Springer Science and Business Media LLC},
   author={Bierlich, Christian and Gustafson, Gösta and Lönnblad, Leif and Shah, Harsh},
   year={2018},
   month=oct }

@misc{pythia8,
      title={A comprehensive guide to the physics and usage of PYTHIA 8.3}, 
      author={Christian Bierlich and Smita Chakraborty and Nishita Desai and Leif Gellersen and Ilkka Helenius and Philip Ilten and Leif Lönnblad and Stephen Mrenna and Stefan Prestel and Christian T. Preuss and Torbjörn Sjöstrand and Peter Skands and Marius Utheim and Rob Verheyen},
      year={2022},
      eprint={2203.11601},
      archivePrefix={arXiv},
      primaryClass={hep-ph},
      url={https://arxiv.org/abs/2203.11601}, 
}

@article{glauber,
   title={Glauber Modeling in High-Energy Nuclear Collisions},
   volume={57},
   ISSN={1545-4134},
   url={http://dx.doi.org/10.1146/annurev.nucl.57.090506.123020},
   DOI={10.1146/annurev.nucl.57.090506.123020},
   number={1},
   journal={Annual Review of Nuclear and Particle Science},
   publisher={Annual Reviews},
   author={Miller, Michael L. and Reygers, Klaus and Sanders, Stephen J. and Steinberg, Peter},
   year={2007},
   month=nov, pages={205–243} 
}

@misc{perlmutter,
    title = "{Perlmutter Architecture}",
    note = "\url{https://docs.nersc.gov/systems/perlmutter/architecture/}"
}

@misc{horovod,
      title={Horovod: fast and easy distributed deep learning in TensorFlow}, 
      author={Alexander Sergeev and Mike Del Balso},
      year={2018},
      eprint={1802.05799},
      archivePrefix={arXiv},
      primaryClass={cs.LG},
      url={https://arxiv.org/abs/1802.05799}, 
}

@misc{tensorflow,
      title={TensorFlow: A system for large-scale machine learning}, 
      author={Martín Abadi and Paul Barham and Jianmin Chen and Zhifeng Chen and Andy Davis and Jeffrey Dean and Matthieu Devin and Sanjay Ghemawat and Geoffrey Irving and Michael Isard and Manjunath Kudlur and Josh Levenberg and Rajat Monga and Sherry Moore and Derek G. Murray and Benoit Steiner and Paul Tucker and Vijay Vasudevan and Pete Warden and Martin Wicke and Yuan Yu and Xiaoqiang Zheng},
      year={2016},
      eprint={1605.08695},
      archivePrefix={arXiv},
      primaryClass={cs.DC},
      url={https://arxiv.org/abs/1605.08695}, 
}

@misc{lion,
      title={Symbolic Discovery of Optimization Algorithms}, 
      author={Xiangning Chen and Chen Liang and Da Huang and Esteban Real and Kaiyuan Wang and Yao Liu and Hieu Pham and Xuanyi Dong and Thang Luong and Cho-Jui Hsieh and Yifeng Lu and Quoc V. Le},
      year={2023},
      eprint={2302.06675},
      archivePrefix={arXiv},
      primaryClass={cs.LG},
      url={https://arxiv.org/abs/2302.06675}, 
}

@article{sp_method,
   title={Methods for analyzing anisotropic flow in relativistic nuclear collisions},
   volume={58},
   ISSN={1089-490X},
   url={http://dx.doi.org/10.1103/PhysRevC.58.1671},
   DOI={10.1103/physrevc.58.1671},
   number={3},
   journal={Physical Review C},
   publisher={American Physical Society (APS)},
   author={Poskanzer, A. M. and Voloshin, S. A.},
   year={1998},
   month=sep, pages={1671–1678} 
}

@article{antikt,
   title={The anti-ktjet clustering algorithm},
   volume={2008},
   ISSN={1029-8479},
   url={http://dx.doi.org/10.1088/1126-6708/2008/04/063},
   DOI={10.1088/1126-6708/2008/04/063},
   number={04},
   journal={Journal of High Energy Physics},
   publisher={Springer Science and Business Media LLC},
   author={Cacciari, Matteo and Salam, Gavin P and Soyez, Gregory},
   year={2008},
   month=apr, pages={063–063} 
}

@article{fastjet,
   title={FastJet user manual: (for version 3.0.2)},
   volume={72},
   ISSN={1434-6052},
   url={http://dx.doi.org/10.1140/epjc/s10052-012-1896-2},
   DOI={10.1140/epjc/s10052-012-1896-2},
   number={3},
   journal={The European Physical Journal C},
   publisher={Springer Science and Business Media LLC},
   author={Cacciari, Matteo and Salam, Gavin P. and Soyez, Gregory},
   year={2012},
   month=mar }

@article{wta,
   title={Jet shapes with the broadening axis},
   volume={2014},
   ISSN={1029-8479},
   url={http://dx.doi.org/10.1007/JHEP04(2014)017},
   DOI={10.1007/jhep04(2014)017},
   number={4},
   journal={Journal of High Energy Physics},
   publisher={Springer Science and Business Media LLC},
   author={Larkoski, Andrew J. and Neill, Duff and Thaler, Jesse},
   year={2014},
   month=apr }

@article{ca_algo,
   title={Better jet clustering algorithms},
   volume={1997},
   ISSN={1029-8479},
   url={http://dx.doi.org/10.1088/1126-6708/1997/08/001},
   DOI={10.1088/1126-6708/1997/08/001},
   number={08},
   journal={Journal of High Energy Physics},
   publisher={Springer Science and Business Media LLC},
   author={Dokshitzer, Yu.L and Leder, G.D and Moretti, S and Webber, B.R},
   year={1997},
   month=aug, pages={001–001} }

@article{softdrop,
   title={Soft drop},
   volume={2014},
   ISSN={1029-8479},
   url={http://dx.doi.org/10.1007/JHEP05(2014)146},
   DOI={10.1007/jhep05(2014)146},
   number={5},
   journal={Journal of High Energy Physics},
   publisher={Springer Science and Business Media LLC},
   author={Larkoski, Andrew J. and Marzani, Simone and Soyez, Gregory and Thaler, Jesse},
   year={2014},
   month=may }

@article{angularities,
   title={Gaining (mutual) information about quark/gluon discrimination},
   volume={2014},
   ISSN={1029-8479},
   url={http://dx.doi.org/10.1007/JHEP11(2014)129},
   DOI={10.1007/jhep11(2014)129},
   number={11},
   journal={Journal of High Energy Physics},
   publisher={Springer Science and Business Media LLC},
   author={Larkoski, Andrew J. and Thaler, Jesse and Waalewijn, Wouter J.},
   year={2014},
   month=nov }

@misc{Zaheer:2017,
      title={Deep Sets}, 
      author={Manzil Zaheer and Satwik Kottur and Siamak Ravanbakhsh and Barnabas Poczos and Ruslan Salakhutdinov and Alexander Smola},
      year={2018},
      eprint={1703.06114},
      archivePrefix={arXiv},
      primaryClass={cs.LG},
      url={https://arxiv.org/abs/1703.06114}, 
}

@article{ampt,
   title={Multiphase transport model for relativistic heavy ion collisions},
   volume={72},
   ISSN={1089-490X},
   url={http://dx.doi.org/10.1103/PhysRevC.72.064901},
   DOI={10.1103/physrevc.72.064901},
   number={6},
   journal={Physical Review C},
   publisher={American Physical Society (APS)},
   author={Lin, Zi-Wei and Ko, Che Ming and Li, Bao-An and Zhang, Bin and Pal, Subrata},
   year={2005},
   month=dec }

@misc{code, 
    title = "{OmniLearn for heavy-ion collisions}", 
    note = "\url{https://github.com/matplo/OmniLearn}" }

\end{document}